\documentclass[prd,aps,showpacs,epsf,floats,onecolumn]{revtex4}%
\usepackage{amssymb}
\usepackage{amsfonts}
\usepackage{amsmath}
\usepackage{graphicx}
\usepackage{fancyhdr}%
\providecommand{\U}[1]{\protect\rule{.1in}{.1in}}
\begin{document}
\title{\textbf{Qubit Geodesics on the Bloch Sphere from Optimal-Speed Hamiltonian
Evolutions}}
\author{\textbf{Carlo Cafaro}$^{1}$ and \textbf{Paul M.\ Alsing}$^{2}$}
\affiliation{$^{1}$SUNY Polytechnic Institute, 12203 Albany, New York, USA}
\affiliation{$^{2}$Air Force Research Laboratory, Information Directorate, 13441 Rome, New
York, USA}

\begin{abstract}
In the geometry of quantum evolutions, a geodesic path is viewed as a path of
minimal statistical length connecting two pure quantum states along which the
maximal number of statistically distinguishable states is minimum. In this
paper, we present an explicit geodesic analysis of the dynamical trajectories
that emerge from the quantum evolution of a single-qubit quantum state. The
evolution is governed by an Hermitian Hamiltonian operator that achieves the
fastest possible unitary evolution between given initial and final pure
states. Furthermore, in addition to viewing geodesics in ray space as paths of
minimal length, we also verify the geodesicity of paths in terms of unit
geometric efficiency and vanishing geometric phase. Finally, based on our
analysis, we briefly address the main hurdles in moving to the geometry of
quantum evolutions for open quantum systems in mixed quantum states.

\end{abstract}

\maketitle

%======== fancyhdr: puts the page number back in on the [R]ight vs the [L]eft =====
\fancyhead[L]{\ifnum\value{page}<2\relax\else\thepage\fi}
%================================================================
%==================

%=================
%needed for fancyhdr
%=================
\thispagestyle{fancy}
%===============

\section{Introduction}

It is well-know that geometry plays a fundamental role in physics. Moreover,
it is not unusual to observe that fundamental physics theories can help
creating new mathematical structures of geometric flavor. For instance, the
noncommutative nature of quantum theory has significantly motivated the birth
of the newly established field of Quantum Riemannian Geometry (QRG,
\cite{beggs20}). This geometry is an extension of classical differential
geometry to curved quantum spaces with a noncommutative coordinate algebra.
For recent applications of QRG to models of quantum gravity and to
formulations of ordinary quantum mechanics in the spirit of classical gravity,
we refer to Refs. \cite{majid19} and \cite{beggs21,beggs22}, respectively.

In this paper, we consider the geometric formulations of quantum mechanics in
a more conventional way where the geometry on the space of quantum states
\cite{karol}, either pure \cite{wootters81} or mixed \cite{braunstein94},
specifies limitations on our capacity of discriminating one state from another
by means of measurements. The geometry on the space of quantum states does not
express, in general, the actual dynamical evolution of a quantum system
\cite{braunstein95}. Indeed, not all Hamiltonian evolutions are shortest time
Hamiltonian evolutions and, therefore, do not coincide with the geodesic paths
on the underlying quantum state space equipped with a suitable metric.
However, focusing for simplicity on pure states, there exist optimum
Hamiltonians generating optimal-speed evolutions characterized by the shortest
duration \cite{brody03,carlini06,brody06,brody07,bender07} along with the
maximal energy dispersion \cite{uhlmann92,ali09}. For such quantum motions,
Hamiltonian curves (i.e., dynamical trajectories) traced by quantum states
undergoing actual physical evolutions can be formally shown to be geodesics
(i.e., geodesic lines or geodesic paths) on the underlying metricized manifolds.

Following Refs. \cite{wootters81,diosi84}, a geodesic path in the above
mentioned geometric formulations of quantum mechanics can be regarded as a
path of minimal statistical length connecting two quantum states along which
the maximal number of statistically distinguishable states is minimum. In
particular, the larger the size of the statistical fluctuations in
measurements prepared to distinguish one state from another, the closer points
are together. Therefore, optimum Hamiltonian evolutions can be shown to happen
along geodesic paths of minimal statistical length. The system evolution
occurs while crossing the minimum number of statistically distinguishable
quantum states and, in addition, moves quickly through regions in which the
energy dispersion is large. The problem of finding a Hamiltonian operator that
achieves the minimum travel time (or, alternatively, the highest evolution
speed) has been considered for systems in either pure
\cite{brody03,carlini06,brody06,brody07,bender07,ali09} or mixed
\cite{carlini08,campaioli19,campbell21,hornedal22} quantum states. In these
works, the emphasis is on the Hamiltonian operator and not on the geodesicity
of the dynamical trajectory traced by the actual Hamiltonian evolution. In
this regard, it is important to keep in mind that geodesics are curves with a
preferred parametrization. Therefore, when characterizing geodesic paths, one
needs to specify its parametrization in terms of a coordinate (that is, the
affine parameter) along with the path (for instance, the great circle for a two-sphere).

The main goal of our paper is to help create more awareness of the interplay
between quantum mechanics and geometry by spelling out how the concept of
shortest time Hamiltonian evolution in the quantum information sense coincides
with the notion of geodesic path in the geometric sense. To achieve this goal,
we consider a shortest time Hermitian Hamiltonian evolution of a two-level
quantum system from a pure source state $\left\vert A\right\rangle $ to a pure
target state $\left\vert B\right\rangle $. Then, we show that the dynamical
trajectory connecting $\left\vert A\right\rangle $ and $\left\vert
B\right\rangle $ that emerges from the unitary evolution operator $U\left(
t\right)  =e^{-\frac{i}{\hslash}\mathrm{H}t}$ coincides with the geodesic path
between the two single-qubit pure states when the two-dimensional Hilbert
space $\mathcal{H}_{2}^{1}$ is viewed in terms of points on the complex
projective Hilbert space $%
%TCIMACRO{\U{2102} }%
%BeginExpansion
\mathbb{C}
%EndExpansion
P^{1}$ (or, equivalently, the Bloch sphere $S^{2}\cong$ $%
%TCIMACRO{\U{2102} }%
%BeginExpansion
\mathbb{C}
%EndExpansion
P^{1}$) equipped with the Fubini-Study metric (or, equivalently, the round
metric on the sphere).

For completeness, we point out that the selection of the main goal of our
paper is also motivated by our recent geometrically oriented investigations on
quantum search algorithms \cite{carlo12,carlo17}, on continuous-time quantum
search evolutions \cite{carlo19A,carlo19B,gassner20}, on the efficiency of
quantum evolutions \cite{carlo20pra,carlo21}, on the emergence of geodesic
paths in different areas of physics (including classical gravity)
\cite{carlo20}, and, finally, on the intriguing link between propagation of
light with maximal degree of coherence and optimal-speed unitary quantum time
evolutions \cite{carlopra22}.

The layout of the rest of the paper is as follows. In Section II, we revisit
some preliminary results on the geometry of pure quantum states. Specifically,
we discuss the concept of distance between pure states, the concept of quantum
line, and the notion of quantum geodesic line with special attention to
suitably chosen parametrizations. In Section III, we present a family of
optimal-speed Hamiltonian evolutions. We describe properties of a typical
Hamiltonian of this family and, in addition, provide an explicit expression of
the shortest time quantum dynamical trajectory connecting an arbitrary initial
source state to an arbitrary final target state during such unitary evolution.
In Section IV, using the geometry of quantum pure states introduced in Section
II and focusing on the quantum Hamiltonian motion described in\ Section III,
we verify in an explicit manner the geodesicity of the quantum dynamical
trajectory emerging from the chosen optimal-speed Hamiltonian evolution. Once
again, we devote special attention to the parametrization of the geodesic
paths. In Section V, we verify geodesicity aspects of the quantum dynamical
trajectory introduced in Section III by means of the concepts of geometric
efficiency \cite{anandan90,carlo20pra} and Berry's geometric phase
\cite{dario04}. In Section VI, we finally present our final remarks.

\section{Geometry of pure quantum states}

In this section, we revisit for completeness some mathematical preliminaries
needed to present our main result. After introducing the concept of
Fubini-Study metric tensor for pure states, our main goals here can be
summarized as follows. First, we discuss general parametrizations of quantum
lines in Eqs. (\ref{geo11}) and (\ref{geo12}). Second, we present quantum
geodesic lines as quantum lines satisfying Eq.(\ref{ass2}) or, equivalently,
as paths of minimal length connecting fixed initial and final states on the
Bloch sphere. For further mathematical details on the geometry of pure quantum
states, we refer to Refs. \cite{provost80,laba17}.

\subsection{Distance between two pure states}

In what follows, we introduce the Fubini-Study metric tensor. Recall that the
finite distance between two quantum states $\left\vert \psi_{A}\right\rangle $
and $\left\vert \psi_{B}\right\rangle $ belonging to a Hilbert space
$\mathcal{H}$ can be defined in different ways. For instance, the Fubini-Study
distance $d_{\text{\textrm{FS}}}\left(  \left\vert \psi_{A}\right\rangle
\text{, }\left\vert \psi_{B}\right\rangle \right)  $ between two quantum
states $\left\vert \psi_{A}\right\rangle $ and $\left\vert \psi_{B}%
\right\rangle $ is defined as \cite{laba17}%
\begin{equation}
d_{\text{\textrm{FS}}}\left(  \left\vert \psi_{A}\right\rangle \text{,
}\left\vert \psi_{B}\right\rangle \right)  \overset{\text{def}}{=}\lambda
\sqrt{1-\left\vert \left\langle \psi_{A}|\psi_{B}\right\rangle \right\vert
^{2}}\text{,} \label{FS}%
\end{equation}
with $\lambda$ being an arbitrary real constant factor. Alternatively, the
Wootters distance $d_{\text{\textrm{W}}}\left(  \left\vert \psi_{A}%
\right\rangle \text{, }\left\vert \psi_{B}\right\rangle \right)  $ between two
quantum states $\left\vert \psi_{A}\right\rangle $ and $\left\vert \psi
_{B}\right\rangle $ is given by \cite{laba17}%
\begin{equation}
d_{\text{\textrm{W}}}\left(  \left\vert \psi_{A}\right\rangle \text{,
}\left\vert \psi_{B}\right\rangle \right)  \overset{\text{def}}{=}\lambda
\cos^{-1}\left[  \left\vert \left\langle \psi_{A}|\psi_{B}\right\rangle
\right\vert \right]  \text{.} \label{W}%
\end{equation}
Interestingly, given two infinitesimally close neighboring pure quantum
states\textbf{ }$\left\vert \psi\left(  \xi\right)  \right\rangle $\textbf{
}and\textbf{ }$\left\vert \psi\left(  \xi+\Delta\xi\right)  \right\rangle
$\textbf{ }that can be distinguished thanks to a real parameter\textbf{ }$\xi
$\textbf{, }it happens that up to the second order in\textbf{ }$\Delta\xi
$\textbf{ }with\textbf{ }$\Delta\xi\ll1$\textbf{, }the differential forms of
the Wootters and the Fubini-Study distances are equivalent
\cite{ravicule97,dodonov99}.

Following the line of reasoning presented in Ref. \cite{laba17}, let us
consider a set of quantum state vectors $\left\{  \left\vert \psi\left(
\xi\right)  \right\rangle \right\}  $ parametrized by the parameters
$\xi\overset{\text{def}}{=}\left(  \xi^{1}\text{,..., }\xi^{m}\right)  $. The
quantity\textbf{ }$m$\textbf{ }denotes the number of real parameters assumed
to parametrize a quantum state\textbf{ }$\left\vert \psi\left(  \xi\right)
\right\rangle $\textbf{ }in\textbf{ }$%
%TCIMACRO{\U{2102} }%
%BeginExpansion
\mathbb{C}
%EndExpansion
P^{n-1}$\textbf{. }For clarity, we assume here that\textbf{ }$\mathcal{H}%
$\textbf{ }is the\textbf{ }$n$-dimensional complex Hilbert space\textbf{
}$\mathcal{H}_{2}^{N}$\textbf{ }of\textbf{ }$N$\textbf{-}qubit quantum states
with\textbf{ }$n=2^{N}$\textbf{ }and we focus on the simple case with
$n=2$\textbf{.} Then, regardless of the chosen definition of finite distance,
the infinitesimal line element $ds_{\text{\textrm{FS}}}^{2}$ quantifying the
distance between two neighboring states $\left\vert \psi\left(  \xi\right)
\right\rangle $ and $\left\vert \psi\left(  \xi+d\xi\right)  \right\rangle $
can be written as%
\begin{equation}
ds_{\text{\textrm{FS}}}^{2}=g_{ab}\left(  \xi\right)  d\xi^{a}d\xi^{b}\text{.}
\label{oggi}%
\end{equation}
The quantity $g_{ab}\left(  \xi\right)  $ in Eq. (\ref{oggi}) is defined as
\cite{provost80},
\begin{equation}
g_{ab}\left(  \xi\right)  \overset{\text{def}}{=}\lambda^{2}\left[
\gamma_{ab}\left(  \xi\right)  -\beta_{a}\left(  \xi\right)  \beta_{b}\left(
\xi\right)  \right]  \text{,} \label{metric}%
\end{equation}
where, given that $\partial_{a}\overset{\text{def}}{=}\partial/\partial\xi
^{a}$, we have%
\begin{equation}
\gamma_{ab}\left(  \xi\right)  \overset{\text{def}}{=}\operatorname{Re}\left[
\left\langle \partial_{a}\psi\left(  \xi\right)  |\partial_{b}\psi\left(
\xi\right)  \right\rangle \right]  \text{, and }\beta_{a}\left(  \xi\right)
\overset{\text{def}}{=}-i\left\langle \psi\left(  \xi\right)  |\partial
_{a}\psi\left(  \xi\right)  \right\rangle \text{.} \label{metrica}%
\end{equation}
Note from Eq. (\ref{metrica}) that,%
\begin{equation}
\beta_{a}\left(  \xi\right)  \beta_{b}\left(  \xi\right)  =\left\langle
\partial_{a}\psi\left(  \xi\right)  |\psi\left(  \xi\right)  \right\rangle
\left\langle \psi\left(  \xi\right)  |\partial_{b}\psi\left(  \xi\right)
\right\rangle \text{,} \label{betaA}%
\end{equation}
since $\partial_{a}\left[  \left\langle \psi\left(  \xi\right)  |\psi\left(
\xi\right)  \right\rangle \right]  =0$ implies that $\left\langle \psi\left(
\xi\right)  |\partial_{a}\psi\left(  \xi\right)  \right\rangle =-\left\langle
\partial_{a}\psi\left(  \xi\right)  |\psi\left(  \xi\right)  \right\rangle $.
Therefore, using Eqs. (\ref{metrica}) and (\ref{betaA}), the metric tensor
components $g_{ab}\left(  \xi\right)  $ defined in Eq. (\ref{metric}) become%
\begin{equation}
g_{ab}\left(  \xi\right)  =\lambda^{2}\left\{  \operatorname{Re}\left[
\left\langle \partial_{a}\psi\left(  \xi\right)  |\partial_{b}\psi\left(
\xi\right)  \right\rangle \right]  -\left\langle \partial_{a}\psi\left(
\xi\right)  |\psi\left(  \xi\right)  \right\rangle \left\langle \psi\left(
\xi\right)  |\partial_{b}\psi\left(  \xi\right)  \right\rangle \right\}
\text{.}%
\end{equation}
For notational simplicity, let us define%
\begin{equation}
A_{ab}\left(  \xi\right)  \overset{\text{def}}{=}\left\langle \partial_{a}%
\psi\left(  \xi\right)  |\psi\left(  \xi\right)  \right\rangle \left\langle
\psi\left(  \xi\right)  |\partial_{b}\psi\left(  \xi\right)  \right\rangle
\text{.} \label{oggi1}%
\end{equation}
We shall prove that $A_{ab}\left(  \xi\right)  d\xi^{a}d\xi^{b}%
=\operatorname{Re}\left[  A_{ab}\left(  \xi\right)  \right]  d\xi^{a}d\xi^{b}%
$. Consider,%
\begin{equation}
A_{ab}\left(  \xi\right)  =\operatorname{Re}\left[  A_{ab}\left(  \xi\right)
\right]  +i\operatorname{Im}\left[  A_{ab}\left(  \xi\right)  \right]
=A_{ab}^{\left(  1\right)  }\left(  \xi\right)  +iA_{ab}^{\left(  2\right)
}\left(  \xi\right)  \text{.}%
\end{equation}
Then, from Eq. (\ref{oggi1}) we have%
\begin{align}
A_{ab}^{\left(  1\right)  }\left(  \xi\right)   &  =\operatorname{Re}\left[
A_{ab}\left(  \xi\right)  \right] \nonumber\\
&  =\operatorname{Re}\left[  \left\langle \partial_{a}\psi\left(  \xi\right)
|\psi\left(  \xi\right)  \right\rangle \left\langle \psi\left(  \xi\right)
|\partial_{b}\psi\left(  \xi\right)  \right\rangle \right] \nonumber\\
&  =\operatorname{Re}\left[  \left\langle \partial_{a}\psi\left(  \xi\right)
|\psi\left(  \xi\right)  \right\rangle ^{\ast}\left\langle \psi\left(
\xi\right)  |\partial_{b}\psi\left(  \xi\right)  \right\rangle ^{\ast}\right]
\nonumber\\
&  =\operatorname{Re}\left[  \left\langle \psi\left(  \xi\right)
|\partial_{a}\psi\left(  \xi\right)  \right\rangle \left\langle \partial
_{b}\psi\left(  \xi\right)  |\psi\left(  \xi\right)  \right\rangle \right]
\nonumber\\
&  =\operatorname{Re}\left[  \left\langle \partial_{b}\psi\left(  \xi\right)
|\psi\left(  \xi\right)  \right\rangle \left\langle \psi\left(  \xi\right)
|\partial_{a}\psi\left(  \xi\right)  \right\rangle \right] \nonumber\\
&  =\operatorname{Re}\left[  A_{ba}\left(  \xi\right)  \right] \nonumber\\
&  =A_{ba}^{\left(  1\right)  }\left(  \xi\right)  \text{,} \label{sym}%
\end{align}
and, in addition,
\begin{align}
A_{ab}^{\left(  2\right)  }\left(  \xi\right)   &  =\operatorname{Im}\left[
A_{ab}\left(  \xi\right)  \right] \nonumber\\
&  =\operatorname{Im}\left[  \left\langle \partial_{a}\psi\left(  \xi\right)
|\psi\left(  \xi\right)  \right\rangle \left\langle \psi\left(  \xi\right)
|\partial_{b}\psi\left(  \xi\right)  \right\rangle \right] \nonumber\\
&  =-\operatorname{Im}\left[  \left\langle \partial_{a}\psi\left(  \xi\right)
|\psi\left(  \xi\right)  \right\rangle ^{\ast}\left\langle \psi\left(
\xi\right)  |\partial_{b}\psi\left(  \xi\right)  \right\rangle ^{\ast}\right]
\nonumber\\
&  =-\operatorname{Im}\left[  \left\langle \partial_{b}\psi\left(  \xi\right)
|\psi\left(  \xi\right)  \right\rangle \left\langle \psi\left(  \xi\right)
|\partial_{a}\psi\left(  \xi\right)  \right\rangle \right] \nonumber\\
&  =-\operatorname{Im}\left[  A_{ba}\left(  \xi\right)  \right] \nonumber\\
&  =-A_{ba}^{\left(  2\right)  }\left(  \xi\right)  \text{.} \label{asym}%
\end{align}
From Eqs. (\ref{sym}) and (\ref{asym}), we conclude that $A_{ab}^{\left(
1\right)  }\left(  \xi\right)  $ is symmetric under exchange of indices while
$A_{ab}^{\left(  2\right)  }\left(  \xi\right)  $ is antisymmetric. Therefore,
from the symmetry of $d\xi^{a}d\xi^{b}$, we have $A_{ab}\left(  \xi\right)
d\xi^{a}d\xi^{b}=\operatorname{Re}\left[  A_{ab}\left(  \xi\right)  \right]
d\xi^{a}d\xi^{b}$. In conclusion, the metric tensor $g_{ab}\left(  \xi\right)
$ can be written as%
\begin{equation}
g_{ab}\left(  \xi\right)  =\lambda^{2}\operatorname{Re}\left[  \left\langle
\partial_{a}\psi\left(  \xi\right)  |\partial_{b}\psi\left(  \xi\right)
\right\rangle \right]  -\lambda^{2}\operatorname{Re}\left[  \left\langle
\partial_{a}\psi\left(  \xi\right)  |\psi\left(  \xi\right)  \right\rangle
\left\langle \psi\left(  \xi\right)  |\partial_{b}\psi\left(  \xi\right)
\right\rangle \right]  \text{,}%
\end{equation}
that is,%
\begin{equation}
g_{ab}\left(  \xi\right)  =\lambda^{2}\operatorname{Re}\left[  \left\langle
\partial_{a}\psi\left(  \xi\right)  |\partial_{b}\psi\left(  \xi\right)
\right\rangle -\left\langle \partial_{a}\psi\left(  \xi\right)  |\psi\left(
\xi\right)  \right\rangle \left\langle \psi\left(  \xi\right)  |\partial
_{b}\psi\left(  \xi\right)  \right\rangle \right]  \text{.} \label{FS1}%
\end{equation}
Eq. (\ref{FS1}) defines the Fubini-Study metric tensor on the manifold of pure
quantum states. We point out that it is convenient to set $\lambda=2$ in\ Eq.
(\ref{FS1}). This way, limiting our attention to the two-dimensional case,
$g_{ab}\left(  \xi\right)  $ in\ Eq. (\ref{FS1}) becomes a metric tensor on
the Bloch sphere with the radius equal to one.

\subsection{Quantum lines}

This subsection is divided in two parts. In the first part, we discuss the
parametrization of quantum lines. In the second part, we show that geodesic
paths are quantum lines of minimal length between two initial and final pure
states on the Bloch sphere. Clearly, the notion of distance used here relies
on the concept of Fubini-Study metric introduced in the previous subsection.

\subsubsection{Parametrization of quantum lines}

Consider two normalized quantum state vectors $\left\vert \psi_{A}%
\right\rangle $ and $\left\vert \psi_{B}\right\rangle $ belonging to a Hilbert
space $\mathcal{H}$ such that,%
\begin{equation}
\left\langle \psi_{A}|\psi_{A}\right\rangle =1\text{, and }\left\langle
\psi_{B}|\psi_{B}\right\rangle =1\text{.}%
\end{equation}
Note that we do not require $\left\vert \psi_{A}\right\rangle $ and
$\left\vert \psi_{B}\right\rangle $ to be orthogonal. Thus, in general,
$\left\langle \psi_{A}|\psi_{B}\right\rangle \neq\delta_{AB}$ with
$\delta_{AB}$ denoting the Kronecker\textbf{ }delta symbol. From $\left\vert
\psi_{A}\right\rangle $ and $\left\vert \psi_{B}\right\rangle $, we can
consider a one-parameter $\xi\in\left(  0\text{, }1\right)  \subset%
%TCIMACRO{\U{211d} }%
%BeginExpansion
\mathbb{R}
%EndExpansion
$ that specifies a parametric set of quantum state vectors $\left\vert
\psi\left(  \xi\right)  \right\rangle $,%
\begin{equation}
\left\vert \psi\left(  \xi\right)  \right\rangle \overset{\text{def}%
}{=}\mathcal{N}_{\xi}\left[  \left(  1-\xi\right)  \left\vert \psi
_{A}\right\rangle +e^{i\varphi}\xi\left\vert \psi_{B}\right\rangle \right]
\text{.} \label{geoline}%
\end{equation}
We point out that $\varphi\in%
%TCIMACRO{\U{211d} }%
%BeginExpansion
\mathbb{R}
%EndExpansion
$ is a relative phase to be properly selected by imposing that global phase
factors are not physically important in quantum mechanics and $\mathcal{N}%
_{\xi}$ is a real normalization factor to be chosen in such a manner that
$\left\langle \psi\left(  \xi\right)  |\psi\left(  \xi\right)  \right\rangle
=1$. Furthermore, we note that $\left\vert \psi\left(  \xi\right)
\right\rangle $ in Eq. (\ref{geoline}) is the analogue of a straight line
$\vec{r}\left(  \xi\right)  $ in a flat Euclidean space that connects two
points $\vec{r}_{A}$ and $\vec{r}_{B}$,%
\begin{equation}
\vec{r}\left(  \xi\right)  \overset{\text{def}}{=}\left(  1-\xi\right)
\vec{r}_{A}+\xi\vec{r}_{B}\text{.} \label{cgeo}%
\end{equation}
Therefore, it appears reasonable to regard the linear combination of the
states $\left\vert \psi_{A}\right\rangle $ and $\left\vert \psi_{B}%
\right\rangle $ that defines $\left\vert \psi\left(  \xi\right)  \right\rangle
$ in Eq. (\ref{geoline}) as a \textquotedblleft geodesic\textquotedblright%
\ line in the Hilbert space $\mathcal{H}$ that connects these two state
vectors. To select the phase $\varphi$, we recall that unlike the classical
case in Eq. (\ref{cgeo}), the quantum case in Eq. (\ref{geoline}) requires
that global phase factors are physically unimportant. Therefore, state vectors
$\left\vert \psi_{j}\right\rangle $ and $e^{i\varphi_{j}}\left\vert \psi
_{j}\right\rangle $ with $j\in\left\{  A\text{, }B\right\}  $ are physically
indistinguishable and represent the same quantum state. For this reason, one
needs to impose that the \textquotedblleft geodesic\textquotedblright\ line
connecting $\left\vert \psi_{A}\right\rangle $ and $\left\vert \psi
_{B}\right\rangle $ must coincide with the \textquotedblleft
geodesic\textquotedblright\ line connecting $\left\vert \tilde{\psi}%
_{A}\right\rangle \overset{\text{def}}{=}e^{i\varphi_{A}}\left\vert \psi
_{A}\right\rangle $ and $\left\vert \tilde{\psi}_{B}\right\rangle
\overset{\text{def}}{=}e^{i\varphi_{B}}\left\vert \psi_{B}\right\rangle $.
Specifically, we require%
\begin{align}
\mathcal{N}_{\xi}\left[  \left(  1-\xi\right)  \left\vert \psi_{A}%
\right\rangle +e^{i\varphi}\xi\left\vert \psi_{B}\right\rangle \right]   &
=\mathcal{N}_{\xi}\left[  \left(  1-\xi\right)  \left\vert \tilde{\psi}%
_{A}\right\rangle +e^{i\tilde{\varphi}}\xi\left\vert \tilde{\psi}%
_{B}\right\rangle \right] \nonumber\\
&  =\mathcal{N}_{\xi}\left[  \left(  1-\xi\right)  e^{i\varphi_{A}}\left\vert
\psi_{A}\right\rangle +e^{i\tilde{\varphi}}\xi e^{i\varphi_{B}}\left\vert
\psi_{B}\right\rangle \right] \nonumber\\
&  =e^{i\varphi_{A}}\mathcal{N}_{\xi}\left[  \left(  1-\xi\right)  \left\vert
\psi_{A}\right\rangle +e^{i\tilde{\varphi}}\xi e^{i\left(  \varphi_{B}%
-\varphi_{A}\right)  }\left\vert \psi_{B}\right\rangle \right] \nonumber\\
&  \sim\mathcal{N}_{\xi}\left[  \left(  1-\xi\right)  \left\vert \psi
_{A}\right\rangle +\xi e^{i\left(  \varphi_{B}-\varphi_{A}\right)  }%
e^{i\tilde{\varphi}}\left\vert \psi_{B}\right\rangle \right]  \text{,}
\label{kk}%
\end{align}
that is,%
\begin{equation}
e^{i\varphi}=e^{i\left(  \varphi_{B}-\varphi_{A}\right)  }e^{i\tilde{\varphi}%
}\text{.} \label{phase}%
\end{equation}
\textbf{ }Observe that in the last line of Eq. (\ref{kk}), the symbol
\textquotedblleft$\sim$\textquotedblright\ denotes physical equivalence of
quantum states and not mathematical equivalence. Eq. (\ref{phase}) can be
satisfied by choosing\textbf{ }the phase factor $e^{i\varphi}$ equal to%
\begin{equation}
e^{i\varphi}=\frac{\left\langle \psi_{B}|\psi_{A}\right\rangle }{\left\vert
\left\langle \psi_{B}|\psi_{A}\right\rangle \right\vert }\text{.}
\label{phase1}%
\end{equation}
Indeed\textbf{,} using Eqs. (\ref{phase}) and (\ref{phase1}), we obtain%
\begin{align}
e^{i\left(  \varphi_{B}-\varphi_{A}\right)  }e^{i\tilde{\varphi}}  &
=e^{i\left(  \varphi_{B}-\varphi_{A}\right)  }\frac{\left\langle \tilde{\psi
}_{B}|\tilde{\psi}_{A}\right\rangle }{\left\vert \left\langle \tilde{\psi}%
_{B}|\tilde{\psi}_{A}\right\rangle \right\vert }\nonumber\\
&  =e^{i\left(  \varphi_{B}-\varphi_{A}\right)  }\frac{e^{-i\left(
\varphi_{B}-\varphi_{A}\right)  }\left\langle \psi_{B}|\psi_{A}\right\rangle
}{\left\vert \left\langle \psi_{B}|\psi_{A}\right\rangle \right\vert
}\nonumber\\
&  =\frac{\left\langle \psi_{B}|\psi_{A}\right\rangle }{\left\vert
\left\langle \psi_{B}|\psi_{A}\right\rangle \right\vert }\nonumber\\
&  =e^{i\varphi}\text{.}%
\end{align}
Therefore, employing the expression of the properly identified phase
factor\textbf{ }$e^{i\varphi}$\textbf{ }in Eq. (\ref{phase1}), the quantum
line in Eq. (\ref{geoline}) can be formally written as%
\begin{equation}
\left\vert \psi\left(  \xi\right)  \right\rangle \overset{\text{def}%
}{=}\mathcal{N}_{\xi}\left[  \left(  1-\xi\right)  \left\vert \psi
_{A}\right\rangle +\left(  \frac{\left\langle \psi_{B}|\psi_{A}\right\rangle
}{\left\vert \left\langle \psi_{B}|\psi_{A}\right\rangle \right\vert }\right)
\xi\left\vert \psi_{B}\right\rangle \right]  \text{.} \label{oggi16}%
\end{equation}
The last quantity that we need to specify in Eq. (\ref{oggi16}) is the real
normalization factor $\mathcal{N}_{\xi}$. As mentioned earlier, this can be
determined by requiring the normalization condition $\left\langle \psi\left(
\xi\right)  |\psi\left(  \xi\right)  \right\rangle =1$. Specifically, we have%
\begin{equation}
1=\left\langle \psi\left(  \xi\right)  |\psi\left(  \xi\right)  \right\rangle
=\mathcal{N}_{\xi}^{2}\left[  1-2\xi\left(  1-\xi\right)  \left(  1-\left\vert
\left\langle \psi_{B}|\psi_{A}\right\rangle \right\vert \right)  \right]
\text{,} \label{normal}%
\end{equation}
that is,
\begin{equation}
\mathcal{N}_{\xi}=\mathcal{N}_{\xi}\left(  \xi\right)  \overset{\text{def}%
}{=}\frac{1}{\sqrt{1-2\xi\left(  1-\xi\right)  \left(  1-\left\vert
\left\langle \psi_{B}|\psi_{A}\right\rangle \right\vert \right)  }}\text{.}
\label{oggi16b}%
\end{equation}
Inserting Eq. (\ref{oggi16b}) into Eq. (\ref{oggi16}), $\left\vert \psi\left(
\xi\right)  \right\rangle $ becomes%
\begin{equation}
\left\vert \psi\left(  \xi\right)  \right\rangle \overset{\text{def}}{=}%
\frac{\left[  \left(  1-\xi\right)  \left\vert \psi_{A}\right\rangle
+\frac{\left\langle \psi_{B}|\psi_{A}\right\rangle }{\left\vert \left\langle
\psi_{B}|\psi_{A}\right\rangle \right\vert }\xi\left\vert \psi_{B}%
\right\rangle \right]  }{\sqrt{1-2\xi\left(  1-\xi\right)  \left(
1-\left\vert \left\langle \psi_{B}|\psi_{A}\right\rangle \right\vert \right)
}}\text{.} \label{geo11}%
\end{equation}
At this point, to explicitly show that $\left\vert \psi\left(  \xi\right)
\right\rangle $ in Eq. (\ref{geo11}) is indeed a proper geodesic line (that
is, a line connecting $\left\vert \psi_{A}\right\rangle $ and $\left\vert
\psi_{B}\right\rangle $ with shortest length with lengths computed by means of
the Fubini-Study metric), it happens to be more convenient employing an
alternative parametrization of the state $\left\vert \psi\left(  \xi\right)
\right\rangle $. This particular step is allowed thanks to the
parametric-invariance of lengths of curves. A convenient parametrization of
$\left\vert \psi\left(  \xi\right)  \right\rangle $ can be given in terms of a
new parameter $\theta\in\left[  0\text{, }\pi\right]  $,%
\begin{equation}
\left\vert \psi\left(  \theta\right)  \right\rangle \overset{\text{def}%
}{=}\mathcal{N}_{\theta}\left[  \cos\left(  \frac{\theta}{2}\right)
\left\vert \psi_{A}\right\rangle +e^{i\varphi}\sin\left(  \frac{\theta}%
{2}\right)  \left\vert \psi_{B}\right\rangle \right]  \text{,} \label{yo1}%
\end{equation}
where the normalization factor $\mathcal{N}_{\theta}$ can be obtained by
imposing the normalization constraint $\left\langle \psi\left(  \theta\right)
|\psi\left(  \theta\right)  \right\rangle =1$. In particular, we have%
\begin{equation}
1=\left\langle \psi\left(  \theta\right)  |\psi\left(  \theta\right)
\right\rangle =\mathcal{N}_{\theta}^{2}\left[  1+\sin\left(  \theta\right)
\left\vert \left\langle \psi_{B}|\psi_{A}\right\rangle \right\vert \right]
\text{,} \label{normal1}%
\end{equation}
that is,
\begin{equation}
\mathcal{N}_{\theta}=\mathcal{N}_{\theta}\left(  \theta\right)
\overset{\text{def}}{=}\frac{1}{\sqrt{1+\sin\left(  \theta\right)  \left\vert
\left\langle \psi_{B}|\psi_{A}\right\rangle \right\vert }}\text{.} \label{yo2}%
\end{equation}
Finally, using Eqs. (\ref{yo1}) and (\ref{yo2}), $\left\vert \psi\left(
\theta\right)  \right\rangle $ becomes%
\begin{equation}
\left\vert \psi\left(  \theta\right)  \right\rangle \overset{\text{def}%
}{=}\frac{\left[  \cos\left(  \frac{\theta}{2}\right)  \left\vert \psi
_{A}\right\rangle +e^{i\varphi}\sin\left(  \frac{\theta}{2}\right)  \left\vert
\psi_{B}\right\rangle \right]  }{\sqrt{1+\sin\left(  \theta\right)  \left\vert
\left\langle \psi_{B}|\psi_{A}\right\rangle \right\vert }}\text{,}
\label{geo12}%
\end{equation}
where $e^{i\varphi}$ in Eq. (\ref{geo12}) equals $\left\langle \psi_{B}%
|\psi_{A}\right\rangle /\left\vert \left\langle \psi_{B}|\psi_{A}\right\rangle
\right\vert $. For completeness, we emphasize that $\left\vert \psi\left(
\xi\right)  \right\rangle $ in\ Eq. (\ref{geo11}) and $\left\vert \psi\left(
\theta\right)  \right\rangle $ in Eq. (\ref{geo12}) are the same states. In
particular, the relation between the two parameters $\xi$ and $\theta$ can be
obtained as follows. From the condition,%
\begin{equation}
\mathcal{N}_{\xi}\left[  \left(  1-\xi\right)  \left\vert \psi_{A}%
\right\rangle +e^{i\varphi}\xi\left\vert \psi_{B}\right\rangle \right]
=\mathcal{N}_{\theta}\left[  \cos\left(  \frac{\theta}{2}\right)  \left\vert
\psi_{A}\right\rangle +e^{i\varphi}\sin\left(  \frac{\theta}{2}\right)
\left\vert \psi_{B}\right\rangle \right]  \text{,}%
\end{equation}
we get%
\begin{equation}
\mathcal{N}_{\xi}\left(  1-\xi\right)  =\mathcal{N}_{\theta}\cos\left(
\frac{\theta}{2}\right)  \text{, and }\mathcal{N}_{\xi}\xi=\mathcal{N}%
_{\theta}\sin\left(  \frac{\theta}{2}\right)  \text{.} \label{this}%
\end{equation}
Manipulations of Eq. (\ref{this}) yield,%
\begin{equation}
\xi=\xi\left(  \theta\right)  \overset{\text{def}}{=}\frac{\tan\left(
\frac{\theta}{2}\right)  }{1+\tan\left(  \frac{\theta}{2}\right)  }\text{.}%
\end{equation}
Observe that $\xi\left(  0\right)  =0$, $\xi\left(  \pi\right)  =1$. Moreover,
for $\xi=0$ and\textbf{ }$\theta=0$, $\left\vert \psi\right\rangle =\left\vert
\psi_{A}\right\rangle $.\textbf{ }Finally, for $\xi=1$ and $\theta=\pi$,
$\left\vert \psi\right\rangle =e^{i\varphi}\left\vert \psi_{B}\right\rangle
\sim\left\vert \psi_{B}\right\rangle $.

\subsubsection{Geodesics as quantum lines of minimal length}

We want to show here that the quantum line in Eq. (\ref{geo12}) is a quantum
geodesic line. For simplicity, let us set $\left\vert \dot{\psi}\right\rangle
=$ $\left\vert \dot{\psi}\left(  \theta\right)  \right\rangle
\overset{\text{def}}{=}\left\vert \partial_{a}\psi\left(  \theta\right)
\right\rangle $ where, in our case, $\partial_{a}=\partial_{\theta
}\overset{\text{def}}{=}\partial/\partial\theta$ (the path depends on a single
parameter). Then, we observe that the single Fubini-Study metric component in
Eq. (\ref{FS1}) can be written in a number of alternative manners%
\begin{align}
g_{\text{\textrm{FS}}}\left(  \theta\right)   &  =\lambda^{2}\operatorname{Re}%
\left[  \left\langle \dot{\psi}|\dot{\psi}\right\rangle -\left\vert
\left\langle \psi|\dot{\psi}\right\rangle \right\vert ^{2}\right] \nonumber\\
&  =\lambda^{2}\operatorname{Re}\left[  \left\langle \dot{\psi}|\dot{\psi
}\right\rangle -\left\langle \psi|\dot{\psi}\right\rangle \left\langle
\psi|\dot{\psi}\right\rangle ^{\ast}\right] \nonumber\\
&  =\lambda^{2}\operatorname{Re}\left[  \left\langle \dot{\psi}|\dot{\psi
}\right\rangle -\left\langle \psi|\dot{\psi}\right\rangle \left\langle
\dot{\psi}|\psi\right\rangle \right] \nonumber\\
&  =\lambda^{2}\operatorname{Re}\left[  \left\langle \dot{\psi}|\dot{\psi
}\right\rangle -\left\langle \dot{\psi}|\psi\right\rangle \left\langle
\psi|\dot{\psi}\right\rangle \right] \nonumber\\
&  =\lambda^{2}\operatorname{Re}\left[  \left\langle \dot{\psi}|\dot{\psi
}\right\rangle +\left\langle \dot{\psi}|\psi\right\rangle \left\langle
\dot{\psi}|\psi\right\rangle \right] \nonumber\\
&  =\lambda^{2}\operatorname{Re}\left[  \left\langle \dot{\psi}|\dot{\psi
}\right\rangle +\left\langle \dot{\psi}|\psi\right\rangle ^{2}\right]
\nonumber\\
&  =\lambda^{2}\operatorname{Re}\left[  \left\langle \dot{\psi}|\dot{\psi
}\right\rangle +\left\langle \psi|\dot{\psi}\right\rangle ^{2}\right]
\text{.} \label{so1}%
\end{align}
In the third to last line of Eq. (\ref{so1}), we have used the fact that
$\left\langle \psi|\psi\right\rangle =1$ implies that $\left\langle \dot{\psi
}|\psi\right\rangle +\left\langle \psi|\dot{\psi}\right\rangle =0$, that is,
$\left\langle \dot{\psi}|\psi\right\rangle =-$ $\left\langle \psi|\dot{\psi
}\right\rangle $.\textbf{ }Therefore\textbf{, }$\operatorname{Re}\left(
\left\langle \psi|\dot{\psi}\right\rangle \right)  =0$ and\textbf{
}$\left\langle \psi|\dot{\psi}\right\rangle $\textbf{ }is a pure imaginary
number\textbf{ }$i\mathcal{A}$\textbf{ }with\textbf{ }$\mathcal{A}\in%
%TCIMACRO{\U{211d} }%
%BeginExpansion
\mathbb{R}
%EndExpansion
$\textbf{. }The real quantity\textbf{ }$\mathcal{A}=\mathcal{A}\left(
t\right)  \overset{\text{def}}{=}-i\left\langle \psi\left(  t\right)
|\dot{\psi}\left(  t\right)  \right\rangle $\textbf{ }is a very relevant
geometric quantity with a significant physical meaning. Indeed, it is
the\textbf{ }connection one form that specifies proper covariant
differentiation and, in addition, leads to the so-called horizontal lift
condition (i.e.,\textbf{ }$i\mathcal{A}\left(  t\right)  =\left\langle
\psi\left(  t\right)  |\dot{\psi}\left(  t\right)  \right\rangle =0$\textbf{)}
\cite{richard}. Moreover, we see later in Eq. (\ref{Berry}) that the
connection one form is such that its line integral gives the geometric phase.
In what follows, we employ the relation%
\begin{equation}
g_{\text{\textrm{FS}}}\left(  \theta\right)  =\lambda^{2}\operatorname{Re}%
\left[  \left\langle \dot{\psi}|\dot{\psi}\right\rangle +\left\langle
\psi|\dot{\psi}\right\rangle ^{2}\right]  \text{.} \label{fs}%
\end{equation}
We want to use Eq. (\ref{geo12}) to evaluate $g_{\text{\textrm{FS}}}\left(
\theta\right)  $ in Eq. (\ref{fs}). We proceed as follows. Note that,%
\begin{equation}
\left\vert \psi\right\rangle =A\left\vert \psi_{A}\right\rangle +e^{i\varphi
}B\left\vert \psi_{B}\right\rangle \text{, and }\left\vert \dot{\psi
}\right\rangle =C\left\vert \psi_{A}\right\rangle +e^{i\varphi}D\left\vert
\psi_{B}\right\rangle \text{,} \label{a}%
\end{equation}
where $A$, $B$, $C$, and $D$ are given by,%
\begin{align}
&  A\overset{\text{def}}{=}\mathcal{N}_{\theta}\cos\left(  \frac{\theta}%
{2}\right)  \text{, }B\overset{\text{def}}{=}\mathcal{N}_{\theta}\sin\left(
\frac{\theta}{2}\right) \nonumber\\
& \nonumber\\
&  C\overset{\text{def}}{=}\mathcal{\dot{N}}_{\theta}\cos\left(  \frac{\theta
}{2}\right)  -\frac{\mathcal{N}_{\theta}}{2}\sin\left(  \frac{\theta}%
{2}\right)  \text{, }D\overset{\text{def}}{=}\mathcal{\dot{N}}_{\theta}%
\sin\left(  \frac{\theta}{2}\right)  +\frac{\mathcal{N}_{\theta}}{2}%
\cos\left(  \frac{\theta}{2}\right) \nonumber\\
& \nonumber\\
&  \mathcal{N}_{\theta}\overset{\text{def}}{=}\frac{1}{\sqrt{1+a\sin\left(
\theta\right)  }}\text{, }a\overset{\text{def}}{=}\left\vert \left\langle
\psi_{B}|\psi_{A}\right\rangle \right\vert \text{, }\mathcal{\dot{N}}_{\theta
}\overset{\text{def}}{=}\frac{-(a/2)\cos\left(  \theta\right)  }{\left[
1+a\sin\left(  \theta\right)  \right]  ^{\frac{3}{2}}}\text{.} \label{a1}%
\end{align}
Inserting Eq. (\ref{a}) into Eq. (\ref{fs}), we obtain%
\begin{equation}
g_{\text{\textrm{FS}}}\left(  \theta\right)  =\lambda^{2}\left\{  C^{2}%
+D^{2}+2aCD+\left[  AC+BD+a\left(  AD+BC\right)  \right]  ^{2}\right\}
\text{.} \label{a2}%
\end{equation}
Then, using Eq. (\ref{a1}) along with performing a number of algebraic
manipulations, $g_{\text{\textrm{FS}}}\left(  \theta\right)  $ in Eq.
(\ref{a2}) becomes%
\begin{equation}
g_{\text{\textrm{FS}}}\left(  \theta\right)  =\frac{\lambda^{2}\left(
1-a^{2}\right)  }{8a\sin\theta-2a^{2}\cos2\theta+2a^{2}+4}\text{,}%
\end{equation}
that is,%
\begin{equation}
g_{\text{\textrm{FS}}}\left(  \theta\right)  =\lambda^{2}\frac{1-a^{2}%
}{4\left[  1+a\sin\left(  \theta\right)  \right]  ^{2}}\text{.}%
\end{equation}
As a side remark, we point out that $\left\langle \psi|\dot{\psi}\right\rangle
=\left[  AC+BD+a\left(  AD+BC\right)  \right]  ^{2}=0$. Indeed, this is
expected since\textbf{ }$i\mathcal{A}\left(  t\right)  =\left\langle
\psi\left(  t\right)  |\dot{\psi}\left(  t\right)  \right\rangle =0$\textbf{
}is the horizontal lift condition that yields geodesics on the Bloch sphere
\cite{richard}. Therefore, the length $s_{A\rightarrow B}$ of the line
connecting the states $\left\vert \psi_{A}\right\rangle $ and $\left\vert
\psi_{B}\right\rangle $ defined as,%
\begin{equation}
s_{A\rightarrow B}\overset{\text{def}}{=}\int_{0}^{\pi}g_{\text{FS}}%
^{1/2}\left(  \theta\right)  d\theta\text{,}%
\end{equation}
is given by,%
\begin{equation}
s_{A\rightarrow B}=\frac{\lambda}{2}\int_{0}^{\pi}\frac{\sqrt{1-a^{2}}%
}{1+a\sin\left(  \theta\right)  }d\theta\text{.}%
\end{equation}
Performing a (Karl Weierstrass) change of variables (that is, $\theta
\rightarrow t=t\left(  \theta\right)  \overset{\text{def}}{=}\tan\left(
\theta/2\right)  $), $s_{A\rightarrow B}$ becomes%
\begin{equation}
s_{A\rightarrow B}=\frac{\lambda}{2}\sqrt{1-a^{2}}\int_{0}^{\infty}\frac
{2}{t^{2}+2at+1}dt\text{.} \label{ass}%
\end{equation}
With the help of the Mathematica symbolic software, we get%
\begin{align}
\int_{0}^{\infty}\frac{2}{t^{2}+2at+1}dt  &  =\frac{2}{\sqrt{1-a^{2}}}\left[
\tan^{-1}\left(  \frac{a+t}{\sqrt{1-a^{2}}}\right)  \right]  _{t=0}^{t=\infty
}\nonumber\\
&  =\frac{2}{\sqrt{1-a^{2}}}\left[  \frac{\pi}{2}-\tan^{-1}\left(  \frac
{a}{\sqrt{1-a^{2}}}\right)  \right] \nonumber\\
&  =\frac{2}{\sqrt{1-a^{2}}}\cos^{-1}\left(  a\right)  \text{.} \label{ass1}%
\end{align}
Finally, using Eq. (\ref{ass1}) and recalling the definition of $a$ in Eq.
(\ref{a1}), the length $s_{A\rightarrow B}$ in Eq. (\ref{ass}) becomes%
\begin{equation}
s_{A\rightarrow B}=\lambda\cos^{-1}\left[  \left\vert \left\langle \psi
_{B}|\psi_{A}\right\rangle \right\vert \right]  \text{.} \label{ass2}%
\end{equation}
Since the length $s_{A\rightarrow B}$ in\ Eq. (\ref{ass2}) of the line
connecting the states $\left\vert \psi_{A}\right\rangle $ and $\left\vert
\psi_{B}\right\rangle $ equals the Wootters distance. For $\lambda=2$, the
Wootters distance equals the angle between the vectors that identify the
initial and final states $\left\vert \psi_{A}\right\rangle $ and $\left\vert
\psi_{B}\right\rangle $ on the Bloch sphere. This angle represents the minimal
possible length of the path $\gamma$ on the Bloch sphere connecting
$\left\vert \psi_{A}\right\rangle $ and $\left\vert \psi_{B}\right\rangle $,%
\begin{equation}
\mathrm{Length}\left(  \gamma_{A\rightarrow B}^{\left(  \mathrm{geodesic}%
\right)  }\right)  \leq\mathrm{Length}\left(  \gamma_{A\rightarrow B}^{\left(
\mathrm{non}\text{-}\mathrm{geodesic}\right)  }\right)  \text{, }
\label{geocondition}%
\end{equation}
for any non-geodesic path $\gamma_{A\rightarrow B}^{\left(  \mathrm{non}%
\text{-}\mathrm{geodesic}\right)  }$. Therefore, we conclude that the quantum
line in Eq. (\ref{geo12}) is indeed a quantum geodesic line.

\section{\textbf{Optimal-Speed Hamiltonian Evolution}}

In this section, having introduced the Fubini-Study metric tensor for pure
states along with the discussion of parametrizations of quantum geodesic paths
as paths of minimal length connecting given initial and final states on the
Bloch sphere, our two main tasks can be stated as follows. First, we introduce
the Hamiltonian operator H in Eq. (\ref{amy}) that achieves the fastest
possible unitary evolution between two given initial and final pure states
$\left\vert A\right\rangle $ and $\left\vert B\right\rangle $.\ Second, we
present the shortest time quantum dynamical trajectory in Eq. (\ref{geodesic})
that emerges from H in Eq. (\ref{amy}) and that connects $\left\vert
A\right\rangle $ to $\left\vert B\right\rangle $.

\subsection{The Hamiltonian}

Following the work presented by Mostafazadeh in Ref. \cite{ali09}, consider a
traceless and time-independent Hamiltonian \textrm{H }specified by a spectral
decomposition given by \textrm{H}$\overset{\text{def}}{=}E_{1}\left\vert
E_{1}\right\rangle \left\langle E_{1}\right\vert +E_{2}\left\vert
E_{2}\right\rangle \left\langle E_{2}\right\vert $, where $\left\langle
E_{2}|E_{1}\right\rangle =\delta_{21}$ and $E_{2}\geq E_{1}$. Clearly,\textbf{
}$\left\{  E_{i}\right\}  _{i=1,2}$\textbf{ }and\textbf{ }$\left\{  \left\vert
E_{i}\right\rangle \right\}  _{i=1,2}$\textbf{ }denote the eigenvalues and the
corresponding orthonormal eigenvectors of the Hamiltonian\textbf{ }\textrm{H}.
Moreover\textbf{, }$\delta_{ij}$\textbf{ }with\textbf{ }$1\leq i$\textbf{,
}$j\leq2$ is the usual Kronecker delta symbol. One is interested in evolving a
state $\left\vert A\right\rangle $, not necessarily normalized, into a state
$\left\vert B\right\rangle $ in the shortest possible time by maximizing the
energy uncertainty $\Delta E$ and obtain $\Delta E=\Delta E_{\max}$, with
\begin{equation}
\Delta E\overset{\text{def}}{=}\left[  \frac{\left\langle A|\mathrm{H}%
^{2}\mathrm{|}A\right\rangle }{\left\langle A|A\right\rangle }-\left(
\frac{\left\langle A|\mathrm{H|}A\right\rangle }{\left\langle A|A\right\rangle
}\right)  ^{2}\right]  ^{1/2}\text{.}%
\end{equation}
We maximize the energy uncertainty\textbf{ }$\Delta E$\textbf{ }since we see
later in Eq. (\ref{relation}) that the speed of quantum evolution
$ds/dt$\textbf{ }along\textbf{ }the curve is proportional to the energy
uncertainty\textbf{ }$\Delta E$\textbf{, }$ds/dt\propto\Delta E$\textbf{.} To
get the value of $\Delta E_{\max}$, we observe that an arbitrary unnormalized
initial state $\left\vert A\right\rangle $ can be recast as $\left\vert
A\right\rangle =\alpha_{1}\left\vert E_{1}\right\rangle +\alpha_{2}\left\vert
E_{2}\right\rangle $ where $\alpha_{1}\overset{\text{def}}{=}$ $\left\langle
E_{1}|A\right\rangle $, $\alpha_{2}\overset{\text{def}}{=}\left\langle
E_{2}|A\right\rangle \in%
%TCIMACRO{\U{2102} }%
%BeginExpansion
\mathbb{C}
%EndExpansion
$. Then, after some straightforward algebra, we obtain%
\begin{equation}
\Delta E=\frac{E_{2}-E_{1}}{2}\left[  1-\left(  \frac{\left\vert \alpha
_{1}\right\vert ^{2}-\left\vert \alpha_{2}\right\vert ^{2}}{\left\vert
\alpha_{1}\right\vert ^{2}+\left\vert \alpha_{2}\right\vert ^{2}}\right)
^{2}\right]  ^{1/2}\text{.} \label{chi4}%
\end{equation}
We observe from Eq. (\ref{chi4}) that the maximum value of $\Delta E$ is
achieved when $\left\vert \alpha_{1}\right\vert =\left\vert \alpha
_{2}\right\vert $ and, in addition, is equal to
\begin{equation}
\Delta E_{\max}\overset{\text{def}}{=}\left(  \frac{E_{2}-E_{1}}{2}\right)
\text{.}%
\end{equation}
A main idea underlying Mostafazadeh's approach in Ref. \cite{ali09} is
expressing \textrm{H}$\overset{\text{def}}{=}E_{1}\left\vert E_{1}%
\right\rangle \left\langle E_{1}\right\vert +E_{2}\left\vert E_{2}%
\right\rangle \left\langle E_{2}\right\vert $ by means of the initial and
final states $\left\vert A\right\rangle $ and $\left\vert B\right\rangle $,
respectively, while keeping $\Delta E=\Delta E_{\max}$. To this end, note that
$\left\vert A\right\rangle $ and $\left\vert B\right\rangle $ can be
decomposed as $\left\vert A\right\rangle =\alpha_{1}\left\vert E_{1}%
\right\rangle +\alpha_{2}\left\vert E_{2}\right\rangle $ and $\left\vert
B\right\rangle =\beta_{1}\left\vert E_{1}\right\rangle +\beta_{2}\left\vert
E_{2}\right\rangle $, respectively. Moreover, we must put $\left\vert
\alpha_{1}\right\vert =\left\vert \alpha_{2}\right\vert $ and $\left\vert
\beta_{1}\right\vert =\left\vert \beta_{2}\right\vert $ to satisfy $\Delta
E=\Delta E_{\max}$ and, consequently, guarantee minimum travel time
$T_{AB}^{\min}$. Therefore, set $\alpha_{2}=e^{i\varphi_{\alpha}}\alpha_{1}$
and $\beta_{2}=e^{i\varphi_{\beta}}\beta_{1}$ with $\varphi_{\alpha}$ and
$\varphi_{\beta}\in%
%TCIMACRO{\U{211d} }%
%BeginExpansion
\mathbb{R}
%EndExpansion
$. Then, states $\left\vert A\right\rangle $ and $\left\vert B\right\rangle $
can be recast as%
\begin{equation}
\left\vert A\right\rangle =\alpha_{1}\left\vert E_{1}\right\rangle +\alpha
_{2}\left\vert E_{2}\right\rangle =\alpha_{1}\left\vert E_{1}\right\rangle
+e^{i\varphi_{\alpha}}\alpha_{1}\left\vert E_{2}\right\rangle \text{,}
\label{chi5a}%
\end{equation}
and,%
\begin{equation}
\left\vert B\right\rangle =\beta_{1}\left\vert E_{1}\right\rangle +\beta
_{2}\left\vert E_{2}\right\rangle =\beta_{1}\left\vert E_{1}\right\rangle
+e^{i\varphi_{\beta}}\beta_{1}\left\vert E_{2}\right\rangle \text{,}
\label{chi5b}%
\end{equation}
respectively. Using Eqs. (\ref{chi5a}) and (\ref{chi5b}), let us introduce the
states $\left\vert \mathcal{A}\right\rangle $ and $\left\vert \mathcal{B}%
\right\rangle $ defined by the relations $\left\vert E_{1}\right\rangle
+e^{i\varphi_{\alpha}}\left\vert E_{2}\right\rangle =\alpha_{1}^{-1}\left\vert
A\right\rangle \overset{\text{def}}{=}\sqrt{2}\left\vert \mathcal{A}%
\right\rangle $ and $\left\vert E_{1}\right\rangle +e^{i\varphi_{\beta}%
}\left\vert E_{2}\right\rangle =\beta_{1}^{-1}\left\vert B\right\rangle
\overset{\text{def}}{=}\sqrt{2}e^{-i\frac{\varphi_{\alpha}-\varphi_{\beta}}%
{2}}\left\vert \mathcal{B}\right\rangle $, respectively. The states\textbf{
}$\left\vert \mathcal{A}\right\rangle $\textbf{ }and\textbf{ }$\left\vert
\mathcal{B}\right\rangle $\textbf{ }are being introduced to express the
Fubini-Study and the geodesic distances in terms of the modulus squared of
their quantum overlap and, in addition, to recast the optimal-speed quantum
Hamiltonian in a convenient form\textbf{. }After some matrix algebra
manipulations with Eqs. (\ref{chi5a}) and (\ref{chi5b}), we obtain%
\begin{equation}
\left(
\begin{array}
[c]{c}%
\left\vert E_{1}\right\rangle \\
\left\vert E_{2}\right\rangle
\end{array}
\right)  =\frac{\sqrt{2}}{e^{i\frac{\varphi_{\alpha}+\varphi_{\beta}}{2}%
}-e^{i\varphi_{\alpha}}e^{i\frac{\varphi_{\alpha}-\varphi_{\beta}}{2}}}\left(
\begin{array}
[c]{cc}%
e^{i\frac{\varphi_{\alpha}+\varphi_{\beta}}{2}} & -e^{i\varphi_{\alpha}}\\
-e^{i\frac{\varphi_{\alpha}-\varphi_{\beta}}{2}} & 1
\end{array}
\right)  \left(
\begin{array}
[c]{c}%
\frac{\alpha_{1}^{-1}}{\sqrt{2}}\left\vert A\right\rangle \\
\frac{\beta_{1}^{-1}}{\sqrt{2}}e^{i\frac{\varphi_{\alpha}-\varphi_{\beta}}{2}%
}\left\vert B\right\rangle
\end{array}
\right)  \text{.} \label{chichi}%
\end{equation}
For completeness, we remark that%
\begin{equation}
\left\vert \left\langle \mathcal{A}|\mathcal{B}\right\rangle \right\vert
^{2}=\frac{\left\vert \left\langle A|B\right\rangle \right\vert ^{2}%
}{\left\langle A|A\right\rangle \left\langle B|B\right\rangle }=\cos
^{2}\left(  \frac{\varphi_{\alpha}-\varphi_{\beta}}{2}\right)  =\cos
^{2}\left(  \frac{\theta_{\mathrm{FS}}}{2}\right)  \text{,}%
\end{equation}
where $\theta_{\mathrm{FS}}\overset{\text{def}}{=}\varphi_{\alpha}%
-\varphi_{\beta}=2s_{\text{\textrm{FS}}}=s_{\text{\textrm{geo}}}$, with
$s_{\text{\textrm{FS}}}$ and $s_{\text{\textrm{geo}}}$ being the Fubini-Study
and the geodesic distances, respectively. Finally, observing that
$E_{2}=-E_{1}\overset{\text{def}}{=}E$ since the Hamiltonian \textrm{H }is
assumed to be traceless and employing Eq. (\ref{chichi}), the spectral
decomposition of the Hamiltonian yields%
\begin{equation}
\mathrm{H}=\frac{iE}{\sin\left(  \frac{\varphi_{\alpha}-\varphi_{\beta}}%
{2}\right)  }\left[  \left\vert \mathcal{B}\right\rangle \left\langle
\mathcal{A}\right\vert -\left\vert \mathcal{A}\right\rangle \left\langle
\mathcal{B}\right\vert \right]  \text{.} \label{Hami}%
\end{equation}
Finally, using Eq. (\ref{chichi}) along with recalling the definitions of
states $\left\vert \mathcal{A}\right\rangle $ and $\left\vert \mathcal{B}%
\right\rangle $, we note that the Hamiltonian in Eq. (\ref{Hami}) can be
expressed in terms of the initial and final states $\left\vert A\right\rangle
$ and $\left\vert B\right\rangle $, respectively, as%

\begin{equation}
\mathrm{H}=iE\cot\left(  \frac{\varphi_{\alpha}-\varphi_{\beta}}{2}\right)
\left[  \frac{\left\vert B\right\rangle \left\langle A\right\vert
}{\left\langle A|B\right\rangle }-\frac{\left\vert A\right\rangle \left\langle
B\right\vert }{\left\langle B|A\right\rangle }\right]  \text{.} \label{amy}%
\end{equation}
Eq.\ (\ref{amy}) describes the correct version of the Hamiltonian specifying
the optimal-speed unitary time evolution as originally proposed in Ref.
\cite{ali09}. For completeness, observe that for $\mathrm{H}$ in Eq.
(\ref{amy}), we correctly get $\left\langle A|\mathrm{H|}A\right\rangle
/\left\langle A|A\right\rangle =0$ and $\Delta E=\left[  \left\langle
A|\mathrm{H}^{2}|A\right\rangle /\left\langle A|A\right\rangle \right]
^{1/2}=E=\Delta E_{\max}$.

\subsection{The quantum dynamical trajectory}

Given the Hamiltonian $\mathrm{H}$ in Eq. (\ref{amy}), we shall find the
quantum dynamical trajectory $t\mapsto\left\vert \psi\left(  t\right)
\right\rangle $ with $\left\vert \psi\left(  t\right)  \right\rangle
\overset{\text{def}}{=}e^{-\frac{i}{\hslash}\mathrm{H}t}\left\vert
A\right\rangle $ connecting initial and final states $\left\vert
A\right\rangle $ and $\left\vert B\right\rangle $, respectively. We shall find
that $\left\vert \psi\left(  t\right)  \right\rangle $ can be written as,%
\begin{equation}
\left\vert \psi\left(  t\right)  \right\rangle =\left[  \cos\left(  \frac
{E}{\hslash}t\right)  -\frac{\cos\left(  \frac{\varphi_{\alpha}-\varphi
_{\beta}}{2}\right)  }{\sin\left(  \frac{\varphi_{\alpha}-\varphi_{\beta}}%
{2}\right)  }\sin\left(  \frac{E}{\hslash}t\right)  \right]  \left\vert
A\right\rangle +\frac{e^{i\frac{\varphi_{\alpha}-\varphi_{\beta}}{2}}}%
{\sin\left(  \frac{\varphi_{\alpha}-\varphi_{\beta}}{2}\right)  }\sin\left(
\frac{E}{\hslash}t\right)  \left\vert B\right\rangle \text{, }
\label{geodesic}%
\end{equation}
where $0\leq t\leq T_{AB}^{\min}$ with $T_{AB}^{\min}\overset{\text{def}%
}{=}\hslash\theta_{\mathrm{FS}}/\left(  2E\right)  $.

First, since we restrict our attention to a traceless Hamiltonian with
$E_{2}=-E_{1}=E$), we note that%
\begin{equation}
\left\vert \psi\left(  t\right)  \right\rangle =e^{-\frac{i}{\hslash
}\mathrm{H}t}\left\vert A\right\rangle =\alpha_{1}e^{\frac{i}{\hslash}%
Et}\left\vert E_{1}\right\rangle +\alpha_{2}e^{-\frac{i}{\hslash}Et}\left\vert
E_{2}\right\rangle \text{.} \label{chi12}%
\end{equation}
Using Eq. (\ref{chichi}), $\left\vert \psi\left(  t\right)  \right\rangle $ in
Eq. (\ref{chi12}) becomes%
\begin{align}
\left\vert \psi\left(  t\right)  \right\rangle  &  =\left[  \frac
{e^{i\frac{\varphi_{\alpha}+\varphi_{\beta}}{2}}e^{\frac{i}{\hslash}Et}%
}{e^{i\frac{\varphi_{\alpha}+\varphi_{\beta}}{2}}-e^{i\varphi_{\alpha}%
}e^{i\frac{\varphi_{\alpha}-\varphi_{\beta}}{2}}}-\frac{e^{i\varphi_{\alpha}%
}e^{i\frac{\varphi_{\alpha}-\varphi_{\beta}}{2}}e^{-\frac{i}{\hslash}Et}%
}{e^{i\frac{\varphi_{\alpha}+\varphi_{\beta}}{2}}-e^{i\varphi_{\alpha}%
}e^{i\frac{\varphi_{\alpha}-\varphi_{\beta}}{2}}}\right]  \left\vert
A\right\rangle +\nonumber\\
& \nonumber\\
&  +\left[  \frac{e^{i\varphi_{\alpha}}e^{i\frac{\varphi_{\alpha}%
-\varphi_{\beta}}{2}}e^{-\frac{i}{\hslash}Et}}{e^{i\frac{\varphi_{\alpha
}+\varphi_{\beta}}{2}}-e^{i\varphi_{\alpha}}e^{i\frac{\varphi_{\alpha}%
-\varphi_{\beta}}{2}}}-\frac{e^{i\varphi_{\alpha}}e^{i\frac{\varphi_{\alpha
}-\varphi_{\beta}}{2}}e^{\frac{i}{\hslash}Et}}{e^{i\frac{\varphi_{\alpha
}+\varphi_{\beta}}{2}}-e^{i\varphi_{\alpha}}e^{i\frac{\varphi_{\alpha}%
-\varphi_{\beta}}{2}}}\right]  \left\vert B\right\rangle \text{.}
\label{chi13}%
\end{align}
After some tedious but straightforward algebra, we note that%
\begin{equation}
\frac{e^{i\frac{\varphi_{\alpha}+\varphi_{\beta}}{2}}}{e^{i\frac
{\varphi_{\alpha}+\varphi_{\beta}}{2}}-e^{i\varphi_{\alpha}}e^{i\frac
{\varphi_{\alpha}-\varphi_{\beta}}{2}}}=-\frac{1}{2i}\frac{e^{-i\frac
{\varphi_{\alpha}-\varphi_{\beta}}{2}}}{\sin\left(  \frac{\varphi_{\alpha
}-\varphi_{\beta}}{2}\right)  }\text{,} \label{chi14}%
\end{equation}
and, in addition,%
\begin{equation}
-\frac{e^{i\varphi_{\alpha}}e^{i\frac{\varphi_{\alpha}-\varphi_{\beta}}{2}}%
}{e^{i\frac{\varphi_{\alpha}+\varphi_{\beta}}{2}}-e^{i\varphi_{\alpha}%
}e^{i\frac{\varphi_{\alpha}-\varphi_{\beta}}{2}}}=\frac{1}{2i}\frac
{e^{i\frac{\varphi_{\alpha}-\varphi_{\beta}}{2}}}{\sin\left(  \frac
{\varphi_{\alpha}-\varphi_{\beta}}{2}\right)  }\text{.} \label{chi15}%
\end{equation}
Therefore, making use of Eqs. (\ref{chi15}) and (\ref{chi14}), Eq.
(\ref{chi13}) yields%
\begin{equation}
\left\vert \psi\left(  t\right)  \right\rangle =\operatorname{Re}\left[
-i\frac{e^{i\frac{\varphi_{\alpha}-\varphi_{\beta}}{2}}}{\sin\left(
\frac{\varphi_{\alpha}-\varphi_{\beta}}{2}\right)  }e^{-\frac{i}{\hslash}%
Et}\right]  \left\vert A\right\rangle +\frac{e^{i\frac{\varphi_{\alpha
}-\varphi_{\beta}}{2}}}{\sin\left(  \frac{\varphi_{\alpha}-\varphi_{\beta}}%
{2}\right)  }\sin\left(  \frac{E}{\hslash}t\right)  \left\vert B\right\rangle
\text{.} \label{chi16}%
\end{equation}
To simplify Eq. (\ref{chi16}), we observe that%
\begin{equation}
\operatorname{Re}\left[  -ie^{i\frac{\varphi_{\alpha}-\varphi_{\beta}}{2}%
}e^{-\frac{i}{\hslash}Et}\right]  =\cos\left(  \frac{E}{\hslash}t\right)
\sin\left(  \frac{\varphi_{\alpha}-\varphi_{\beta}}{2}\right)  -\sin\left(
\frac{E}{\hslash}t\right)  \cos\left(  \frac{\varphi_{\alpha}-\varphi_{\beta}%
}{2}\right)  \text{.} \label{chi17}%
\end{equation}
Finally, using Eq. (\ref{chi17}), $\left\vert \psi\left(  t\right)
\right\rangle $ in Eq. (\ref{chi16}) becomes $\left\vert \psi\left(  t\right)
\right\rangle $ in Eq. (\ref{geodesic}). For completeness, observe that%
\begin{equation}
\left\vert \psi\left(  0\right)  \right\rangle =\left\vert A\right\rangle
\text{, and }\left\vert \psi\left(  \frac{\hslash}{E}\frac{\varphi_{\alpha
}-\varphi_{\beta}}{2}\right)  \right\rangle =e^{i\frac{\varphi_{\alpha
}-\varphi_{\beta}}{2}}\left\vert B\right\rangle \simeq\left\vert
B\right\rangle \text{.}%
\end{equation}
Moreover, as a consistency check, observe that $\left\langle \psi\left(
t\right)  |\psi\left(  t\right)  \right\rangle =1$ where $\left\langle
A|B\right\rangle =\exp\left(  -i\frac{\varphi_{\alpha}-\varphi_{\beta}}%
{2}\right)  \cos\left[  \left(  \varphi_{\alpha}-\varphi_{\beta}\right)
/2\right]  $.

\section{Geodesicity of the quantum dynamical trajectory}

\begin{table}[t]
\centering
\begin{tabular}
[c]{c|c|c}\hline\hline
\textbf{Sequence of steps} & \textbf{Temporal parametrizations} &
\textbf{Quantum states}\\\hline
Step-1 & $t$, with $0\leq t\leq\hslash\theta_{\mathrm{FS}}/\left(  2E\right)
$ & $\left\vert \psi\left(  t\right)  \right\rangle $\\
Step-2 & $\xi$, with $0\leq\xi\leq1$ & $\left\vert \psi_{\mathrm{geo}}\left(
\xi\right)  \right\rangle $\\
Step-3 & $\eta$, with $0\leq\eta\leq\pi$ & $\left\vert \tilde{\psi
}_{\mathrm{geo}}\left(  \eta\right)  \right\rangle $\\\hline
\end{tabular}
\caption{Schematic description of the steps performed to prove the geodesicity
of the path traced by the pure state vector subjected to the chosen quantum
evolution. Each step is characterized by a temporal parameter and a
corresponding parametrized state vector. The transition from $\left\vert
\psi\left(  t\right)  \right\rangle $ to $\left\vert \tilde{\psi
}_{\mathrm{geo}}\left(  \eta\right)  \right\rangle $ allowed us to explicitly
verify the geodesic nature of the path traced by the state vector $\left\vert
\tilde{\psi}_{\mathrm{geo}}\left(  \eta\right)  \right\rangle $.}%
\end{table}

In this section, exploiting geometric tools to describe pure states as
presented in Section II and focusing on the Hamiltonian motion specified
in\ Section III, we explicitly show the geodesicity (as defined in Eq.
(\ref{ass2})) of the shortest time quantum dynamical trajectory in Eq.
(\ref{geodesic}) that emerges from the chosen optimal-speed Hamiltonian
evolution H in\ Eq. (\ref{amy}). While doing so, we devote special attention
to the parametrization of quantum geodesic paths.

We begin by using the formalism presented in Section II to show that
$\left\vert \psi\left(  t\right)  \right\rangle $ in Eq. (\ref{geodesic})
defines a geodesic path on the Bloch sphere. We start by performing a sequence
of two changes of parametrization of the vector state $\left\vert \psi\left(
t\right)  \right\rangle $ in Eq. (\ref{geodesic}) with $0\leq t\leq\left(
\hslash/E\right)  \left[  \left(  \varphi_{\alpha}-\varphi_{\beta}\right)
/2\right]  =\hslash\theta_{\mathrm{FS}}/\left(  2E\right)  $. In our first
reparametrization, we recast $\left\vert \psi\left(  t\right)  \right\rangle $
as $\left\vert \psi_{\text{\textrm{geo}}}\left(  \xi\right)  \right\rangle $
given by%
\begin{equation}
\left\vert \psi\left(  t\right)  \right\rangle =\left\vert \psi
_{\text{\textrm{geo}}}\left(  \xi\left(  t\right)  \right)  \right\rangle
\text{,}%
\end{equation}
where,%
\begin{equation}
\left\vert \psi_{\text{\textrm{geo}}}\left(  \xi\right)  \right\rangle
\overset{\text{def}}{=}\mathcal{N}_{\xi}\left(  \xi\right)  \left[  \left(
1-\xi\right)  \left\vert A\right\rangle +\xi e^{i\phi}\left\vert
B\right\rangle \right]  \text{,} \label{GG23}%
\end{equation}
with $0\leq\xi\leq1$. Recall that the normalization factor $\mathcal{N}_{\xi
}\left(  \xi\right)  $ and the phase factor $e^{i\phi}$ in Eq. (\ref{GG23})
are given by
\begin{equation}
\mathcal{N}_{\xi}\left(  \xi\right)  \overset{\text{def}}{=}\frac{1}%
{\sqrt{1-2\xi\left(  1-\xi\right)  \left(  1-\left\vert \left\langle
A|B\right\rangle \right\vert \right)  }}\text{, and }e^{i\phi}%
\overset{\text{def}}{=}\frac{\left\langle B|A\right\rangle }{\left\vert
\left\langle B|A\right\rangle \right\vert }\text{,} \label{GG24}%
\end{equation}
respectively. Clearly, we have to find the expression of $\xi=\xi\left(
t\right)  $. In our case, note that%
\begin{equation}
e^{i\phi}\overset{\text{def}}{=}\frac{\left\langle B|A\right\rangle
}{\left\vert \left\langle B|A\right\rangle \right\vert }=\frac{\left\langle
A|B\right\rangle ^{\ast}}{\left\vert \left\langle B|A\right\rangle \right\vert
}=\frac{e^{i\frac{\varphi_{\alpha}-\varphi_{\beta}}{2}}\cos\left(
\frac{\varphi_{\alpha}-\varphi_{\beta}}{2}\right)  }{\left\vert e^{i\frac
{\varphi_{\alpha}-\varphi_{\beta}}{2}}\cos\left(  \frac{\varphi_{\alpha
}-\varphi_{\beta}}{2}\right)  \right\vert }=e^{i\frac{\varphi_{\alpha}%
-\varphi_{\beta}}{2}}\text{.}%
\end{equation}
Therefore, to recast $\left\vert \psi\left(  t\right)  \right\rangle $ as
$\left\vert \psi_{\text{\textrm{geo}}}\left(  \xi\left(  t\right)  \right)
\right\rangle $, we need to solve the following algebraic system of equations%
\begin{equation}
\left\{
\begin{array}
[c]{c}%
\mathcal{N}_{\xi}\left(  \xi\right)  \xi=\frac{\sin\left(  \frac{E}{\hslash
}t\right)  }{\sin\left(  \frac{\varphi_{\alpha}-\varphi_{\beta}}{2}\right)
}\text{,}\\
\\
\mathcal{N}_{\xi}\left(  \xi\right)  \left(  1-\xi\right)  =\cos\left(
\frac{E}{\hslash}t\right)  -\frac{\cos\left(  \frac{\varphi_{\alpha}%
-\varphi_{\beta}}{2}\right)  }{\sin\left(  \frac{\varphi_{\alpha}%
-\varphi_{\beta}}{2}\right)  }\sin\left(  \frac{E}{\hslash}t\right)  \text{.}%
\end{array}
\right.  \label{SS1}%
\end{equation}
After some algebra, we get from Eq. (\ref{SS1}) that
\begin{equation}
\xi\left(  t\right)  =\frac{\sin\left(  \frac{E}{\hslash}t\right)  }%
{\cos\left(  \frac{E}{\hslash}t\right)  \sin\left(  \frac{\varphi_{\alpha
}-\varphi_{\beta}}{2}\right)  +\left[  1-\cos\left(  \frac{\varphi_{\alpha
}-\varphi_{\beta}}{2}\right)  \right]  \sin\left(  \frac{E}{\hslash}t\right)
}\text{.} \label{eta}%
\end{equation}
For completeness, we remark that we correctly obtain from the relation in Eq.
(\ref{eta}) that%
\begin{equation}
\xi\left(  t\right)  =0\text{, and }\xi\left(  \frac{\hslash}{E}\frac
{\varphi_{\alpha}-\varphi_{\beta}}{2}\right)  =1\text{.}%
\end{equation}
We also find that the normalization factor $\mathcal{N}_{\xi}\left(
\xi\right)  $ in Eq. (\ref{GG24}) becomes%
\begin{equation}
\mathcal{N}_{\xi}\left(  \xi\right)  =\cos\left(  \frac{E}{\hslash}t\right)
+\frac{1-\cos\left(  \frac{\varphi_{\alpha}-\varphi_{\beta}}{2}\right)  }%
{\sin\left(  \frac{\varphi_{\alpha}-\varphi_{\beta}}{2}\right)  }\sin\left(
\frac{E}{\hslash}t\right)  \text{.}%
\end{equation}
In our second reparametrization, we recast $\left\vert \psi
_{\text{\textrm{geo}}}\left(  \xi\right)  \right\rangle $ in Eq. (\ref{GG23})
as%
\begin{equation}
\left\vert \psi_{\text{\textrm{geo}}}\left(  \xi\right)  \right\rangle
=\left\vert \tilde{\psi}_{\text{\textrm{geo}}}\left(  \eta\left(  \xi\right)
\right)  \right\rangle
\end{equation}
where,%
\begin{equation}
\left\vert \tilde{\psi}_{\text{\textrm{geo}}}\left(  \eta\right)
\right\rangle \overset{\text{def}}{=}\mathcal{N}_{\eta}\left(  \eta\right)
\left[  \cos\left(  \frac{\eta}{2}\right)  \left\vert A\right\rangle
+\sin\left(  \frac{\eta}{2}\right)  e^{i\phi}\left\vert B\right\rangle
\right]  \text{,} \label{para2}%
\end{equation}
with $0\leq\eta\leq\pi$. Within this new parametrization, the normalization
factor $\mathcal{N}_{\eta}\left(  \eta\right)  $ and the phase factor
$e^{i\phi}$ in Eq. (\ref{para2}) are given by
\begin{equation}
\mathcal{N}_{\eta}\left(  \eta\right)  \overset{\text{def}}{=}\frac{1}%
{\sqrt{1+\sin\left(  \eta\right)  \left\vert \left\langle B|A\right\rangle
\right\vert }}\text{, and }e^{i\phi}\overset{\text{def}}{=}\frac{\left\langle
B|A\right\rangle }{\left\vert \left\langle B|A\right\rangle \right\vert
}\text{,}%
\end{equation}
respectively. In particular, we find
\begin{equation}
\xi\left(  \eta\right)  =\frac{\tan\left(  \frac{\eta}{2}\right)  }%
{1+\tan\left(  \frac{\eta}{2}\right)  }\text{,}%
\end{equation}
that is,%
\begin{equation}
\eta\left(  \xi\right)  =2\tan^{-1}\left(  \frac{\xi}{1-\xi}\right)  \text{,}%
\end{equation}
with $0\leq\eta\left(  \xi\right)  \leq\pi$. In summary, $\left\vert
\psi\left(  t\right)  \right\rangle $ in Eq. (\ref{geodesic}) can be recast as%
\begin{equation}
\left\vert \psi\left(  t\right)  \right\rangle =\left\vert \psi
_{\text{\textrm{geo}}}\left(  \xi\left(  t\right)  \right)  \right\rangle
=\left\vert \tilde{\psi}_{\text{\textrm{geo}}}\left(  \eta\left(  \xi\left(
t\right)  \right)  \right)  \right\rangle \text{,} \label{impogeo}%
\end{equation}
where,%
\begin{equation}
\eta\left(  \xi\left(  t\right)  \right)  \overset{\text{def}}{=}2\tan
^{-1}\left(  \frac{\xi\left(  t\right)  }{1-\xi\left(  t\right)  }\right)
\text{,} \label{chissa2}%
\end{equation}
with $\xi\left(  t\right)  $ given in Eq. (\ref{eta}). Substituting Eq.
(\ref{eta}) into Eq. (\ref{chissa2}), we obtain%
\begin{equation}
\eta\left(  t\right)  \overset{\text{def}}{=}2\tan^{-1}\left[  \frac
{\sin\left(  \frac{E}{\hslash}t\right)  }{\sin\left(  \frac{\theta
_{\mathrm{FS}}}{2}\right)  \cos\left(  \frac{E}{\hslash}t\right)  -\cos\left(
\frac{\theta_{\mathrm{FS}}}{2}\right)  \sin\left(  \frac{E}{\hslash}t\right)
}\right]  \text{,} \label{ggg}%
\end{equation}
with $0\leq t\leq\frac{\hslash\theta_{\mathrm{FS}}}{2E}$. For consistency
check, note that we correctly have $\eta\left(  0\right)  =0$ and $\eta\left(
t\right)  \rightarrow\pi$ as $t\rightarrow\left(  \hslash\theta_{\mathrm{FS}%
}\right)  /2E$ since
\begin{equation}
\lim_{t\rightarrow\left(  \frac{\hslash\theta_{\mathrm{FS}}}{2E}\right)  ^{-}%
}\frac{\sin\left(  \frac{E}{\hslash}t\right)  }{\sin\left(  \frac{\theta}%
{2}\right)  \cos\left(  \frac{E}{\hslash}t\right)  -\cos\left(  \frac{\theta
}{2}\right)  \sin\left(  \frac{E}{\hslash}t\right)  }=+\infty\text{.}%
\end{equation}
Having recast $\left\vert \psi\left(  t\right)  \right\rangle $ in Eq.
(\ref{geodesic}) as $\left\vert \tilde{\psi}_{\text{\textrm{geo}}}\left(
\eta\right)  \right\rangle $ in Eq. (\ref{para2}) with $\eta=\eta\left(
t\right)  $ in Eq. (\ref{ggg}), we can follow the analysis outlined in the
last part of Section II to verify that the distance of the path traced out by
$t\mapsto\left\vert \psi\left(  t\right)  \right\rangle $ between $\left\vert
A\right\rangle $ and $\left\vert B\right\rangle $ on the Bloch sphere is equal
to the Wootters distance. Therefore, we can conclude that the Hamiltonian
\textrm{H }in Eq. (\ref{amy}) gives rise to a trajectory $t\mapsto\left\vert
\psi\left(  t\right)  \right\rangle $ that represents a geodesic line on the
Bloch sphere. This concludes our quantitative discussion. However, before
presenting our final remarks, we briefly present two alternative ways to check
the geodesicity of a curve in ray space in the next section.\begin{figure}[t]
\centering
\includegraphics[width=0.35\textwidth] {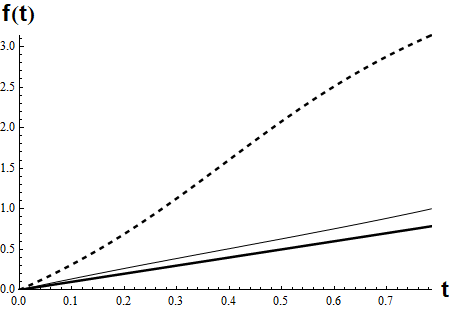}\caption{Plots of the strictly
monotonic functions $f_{1}\left(  t\right)  \protect\overset{\text{def}}{=}t$
(thick solid line), $f_{2}\left(  t\right)  \protect\overset{\text{def}}{=}%
\xi\left(  t\right)  $ (thin solid line), and $f_{3}\left(  t\right)
\protect\overset{\text{def}}{=}\eta\left(  t\right)  $ (dashed line) versus
$t$ with $0\leq t\leq\pi/4$. For simplicity, we set in the plot $E=1$,
$\hslash=1$, and $\left(  \varphi_{\alpha}-\varphi_{\beta}\right)  /2=\pi/4$.
The geodesicity of the path traced by the optimally evolved quantum states
$\left\vert \psi\left(  t\right)  \right\rangle $ was checked in terms of the
state parametrized as $\left\vert \tilde{\psi}_{\mathrm{geo}}\left(
\eta\right)  \right\rangle $ with $\left\vert \psi\left(  t\right)
\right\rangle \protect\overset{\text{def}}{=}\left\vert \tilde{\psi
}_{\mathrm{geo}}\left(  \eta\left(  \xi\left(  t\right)  \right)  \right)
\right\rangle $.}%
\end{figure}

\section{Alternative consistency checks of the geodesicity of a curve}

In the previous section, we have verified in an explicit manner the
geodesicity of the curve in projective Hilbert space emerging from the
optimal-speed Hamiltonian in\ Eq. (\ref{amy}) by showing that the curve in Eq.
(\ref{geodesic}) is a minimal length curve (Eq. (\ref{ass2})). In this
section, we check the geodesicity property in two additional manners. In the
first verification, specified by a necessary and sufficient criterion, we
observe that the curve in ray space is a unit geometric efficiency curve
\cite{anandan90}. In the second verification, which provides a necessary but
not sufficient criterion, we check that the curve in ray space is a null phase
curve (i.e., the geometric phase vanishes \cite{karol,mukunda93,mittal22}).

\subsection{Geodesics as unit geometric efficiency curves}

We begin by discussing a geometric measure of efficiency for a quantum
evolution \cite{anandan90}. Consider an evolution of a state vector
$\left\vert \psi\left(  t\right)  \right\rangle $ specified \ by the
Schr\"{o}dinger equation, $i\hslash\partial_{t}\left\vert \psi\left(
t\right)  \right\rangle =\mathrm{H}\left(  t\right)  \left\vert \psi\left(
t\right)  \right\rangle $, with $t_{A}\leq t\leq t_{B}$. Then, following
\cite{anandan90}, a geometric measure of efficiency $\eta_{\mathrm{geometric}%
}$ with $0\leq\eta_{\mathrm{geometric}}\leq1$ for such a quantum evolution is
given by \cite{carlo20pra}%
\begin{equation}
\eta_{\mathrm{geometric}}\overset{\text{def}}{=}1-\frac{\Delta s}{s}%
=\frac{2\cos^{-1}\left[  \left\vert \left\langle A|B\right\rangle \right\vert
\right]  }{2\int_{t_{A}}^{t_{B}}\frac{\Delta E\left(  t^{\prime}\right)
}{\hslash}dt^{\prime}}\text{,} \label{efficiency}%
\end{equation}
where $\Delta s\overset{\text{def}}{=}s-s_{0}$, $s_{0}$ is the distance along
the shortest geodesic path joining the distinct initial $\left\vert
A\right\rangle \overset{\text{def}}{=}$ $\left\vert \psi\left(  t_{A}\right)
\right\rangle $ and final $\left\vert B\right\rangle \overset{\text{def}%
}{=}\left\vert \psi\left(  t_{B}\right)  \right\rangle $ states on the
projective Hilbert space and finally, $s$ denotes the distance along the
dynamical trajectory traced by the state vector $\left\vert \psi\left(
t\right)  \right\rangle $ with $t_{A}\leq t\leq t_{B}$. Clearly, a geodesic
quantum evolution is specified by the condition%
\begin{equation}
\eta_{\mathrm{geometric}}^{\left(  \gamma_{\mathrm{geodesic}}\right)
}=1\text{.} \label{unitefficiency}%
\end{equation}
Note that the numerator in\ Eq. (\ref{efficiency}) specifies the angle between
the state vectors $\left\vert A\right\rangle $ and $\left\vert B\right\rangle
$ and is equal to the Wootters distance \cite{wootters81}. Instead, the
denominator in Eq. (\ref{efficiency}) describes the integral of the
infinitesimal distance $ds$ along the evolution curve in ray space
\cite{anandan90}, $ds\overset{\text{def}}{=}2\left[  \Delta E\left(  t\right)
/\hslash\right]  dt$, with $\Delta E\overset{\text{def}}{=}\left[
\left\langle \psi|\mathrm{H}^{2}\left(  t\right)  |\psi\right\rangle
-\left\langle \psi|\mathrm{H}\left(  t\right)  |\psi\right\rangle ^{2}\right]
^{1/2}$ denoting the square root of the dispersion of the Hamiltonian operator
$\mathrm{H}\left(  t\right)  $. Remarkably, Anandan and Aharonov demonstrated
that the infinitesimal distance $ds\overset{\text{def}}{=}2\left[  \Delta
E\left(  t\right)  /\hslash\right]  dt$ is linked to the Fubini-Study
infinitesimal distance $ds_{\text{\textrm{FS}}}$ \cite{anandan90},%
\begin{equation}
ds_{\text{\textrm{FS}}}^{2}\left(  \left\vert \psi\left(  t\right)
\right\rangle \text{, }\left\vert \psi\left(  t+dt\right)  \right\rangle
\right)  \overset{\text{def}}{=}4\left[  1-\left\vert \left\langle \psi\left(
t\right)  |\psi\left(  t+dt\right)  \right\rangle \right\vert ^{2}\right]
=4\frac{\Delta E^{2}\left(  t\right)  }{\hslash^{2}}dt^{2}+\mathcal{O}\left(
dt^{3}\right)  \text{,} \label{relation}%
\end{equation}
where $\mathcal{O}\left(  dt^{3}\right)  $ denotes an infinitesimal quantity
equal or higher than $dt^{3}$. From the link between $ds_{\mathrm{FS}}$ and
$ds$, one concludes that $s$ is proportional to the temporal integral of the
energy uncertainty$\Delta E$ of the quantum system and, in addition, specifies
the distance along the quantum evolution of the system in ray space as
measured by the Fubini-Study metric. We emphasize that $\Delta s$ is equal to
zero and the efficiency $\eta$ in Eq. (\ref{efficiency}) reduces to one when
the dynamical curve coincides with the shortest geodesic path joining the
initial and final states. Obviously, the shortest possible distance between
two orthogonal quantum states in ray space is $\pi$.

In our problem, $\left\langle A|B\right\rangle \overset{\text{def}%
}{=}e^{-i\frac{\varphi_{\alpha}-\varphi_{\beta}}{2}}\cos\left[  \left(
\varphi_{\alpha}-\varphi_{\beta}\right)  /2\right]  $, $\Delta E\left(
t^{\prime}\right)  \overset{\text{def}}{=}E=$\textrm{const}., $t_{A}%
\overset{\text{def}}{=}0$, and $t_{B}\overset{\text{def}}{=}\left(
\hslash/E\right)  \left[  \left(  \varphi_{\alpha}-\varphi_{\beta}\right)
/2\right]  $. Therefore, a simple calculation yields a unit geometric
efficiency in Eq. (\ref{efficiency}), $\eta=1$. Therefore, the geodesicity
condition is properly satisfied.

\subsection{Geodesics as null phase curves}

In this second subsection, we check that the curve in ray space is a null
phase curve \cite{karol,mukunda93,mittal22}). We shall check this condition by
showing that the total phase along the horizontal lift of a geodesic in the
projective Hilbert space is zero. This is a necessary (but not sufficient;
there are null phase curves that are not necessarily geodesic curves)
condition to be satisfied by a geodesic curve in ray space as recently
stressed in Refs. \cite{mittal22}. Before presenting the simple check, we
provide some basic mathematical background along with some relevant historical
remarks on the concept of geometric phase.

\emph{Basic background}.\emph{ }Let $\mathcal{H}\backslash\left\{  0\right\}
$ denote the Hilbert space described by an $\left(  N+1\right)  $-dimensional
complex vector space of normalized state vectors $\left\{  \left\vert
\psi\left(  t\right)  \right\rangle \right\}  $. In quantum mechanics, a
physical state is not represented by a normalized state vector $\left\vert
\psi\left(  t\right)  \right\rangle \in\mathcal{H}\backslash\left\{
0\right\}  $. Instead, physical states are represented by a ray.\ A ray is the
one-dimensional subspace $\left\{  e^{i\phi\left(  t\right)  }\left\vert
\psi\left(  t\right)  \right\rangle :e^{i\phi\left(  t\right)  }\in
U(1)\right\}  $ of $\mathcal{H}\backslash\left\{  0\right\}  $ to which this
vector $\left\vert \psi\left(  t\right)  \right\rangle $ belongs. Two state
vectors $\left\vert \psi_{1}\left(  t\right)  \right\rangle $ and $\left\vert
\psi_{2}\left(  t\right)  \right\rangle $ that belong to the same ray are
equivalent, $\left\vert \psi_{1}\left(  t\right)  \right\rangle \sim\left\vert
\psi_{2}\left(  t\right)  \right\rangle $, if $\left\vert \psi_{1}\left(
t\right)  \right\rangle =e^{i\phi_{12}\left(  t\right)  }\left\vert \psi
_{2}\left(  t\right)  \right\rangle $ for some $\phi_{12}\left(  t\right)  \in%
%TCIMACRO{\U{211d} }%
%BeginExpansion
\mathbb{R}
%EndExpansion
$. The equivalence relation \textquotedblleft$\sim$\textquotedblright\ gives
rise to equivalence classes on the $\left(  2N+1\right)  $-dimensional sphere
$S^{2N+1}$. The set of all equivalence classes $S^{2N+1}/U\left(  1\right)  $
determines the space of rays (that is, the space of physical states). In
general, $S^{2N+1}/U\left(  1\right)  $ is called the projective Hilbert space
$\mathcal{P}\left(  \mathcal{H}\right)  $.

This relation between state vectors in Hilbert space and rays in projective
Hilbert space mediated by phase factors can be nicely described in terms of
the fiber bundle formalism \cite{karol,nakahara}. Roughly speaking, the main
ingredients of a fiber bundle are a total space $E$, a base space $M$, a fiber
space $\mathrm{F}$, a group $G$ acting on the fibers, and \ a projection map
$\pi$ that projects the fibers above to points in $M$. In quantum mechanics,
$\mathcal{H}\backslash\left\{  0\right\}  $ plays the role of $E$,
$\mathcal{P}\left(  \mathcal{H}\right)  $ plays the part of $M$, $U\left(
1\right)  $ plays the role of $G$, fibers in $F$ are represented by all unit
vectors from the same ray and, finally, the projection map $\pi$ given by%
\begin{equation}
\pi:\mathcal{H}\backslash\left\{  0\right\}  \ni\left\vert \psi\left(
t\right)  \right\rangle \mapsto\pi\left(  \left\vert \psi\left(  t\right)
\right\rangle \right)  \overset{\text{def}}{=}\left\vert \psi\left(  t\right)
\right\rangle \left\langle \psi\left(  t\right)  \right\vert \in
\mathcal{P}\left(  \mathcal{H}\right)  \text{,}%
\end{equation}
plays the part of the projection in the fiber bundle construction. For more
details on the fiber bundle formalism, we refer to Refs.
\cite{nakahara,eguchi80,bohm91}. A schematic summary of the fiber bundle
formalism in quantum mechanics appears in Table II.\begin{table}[t]
\centering
\begin{tabular}
[c]{c|c|c}\hline\hline
\textbf{Fiber bundle ingredients} & \textbf{Symbols} & \textbf{Elements}%
\\\hline
Total space & $\mathcal{H}\backslash\left\{  0\right\}  $ & State vectors\\
Base space & $\mathcal{P}\left(  \mathcal{H}\right)  $ & Rays\\
Fiber space & $\mathrm{F}$ & Fibers\\
Structure group & $U\left(  1\right)  $ & Phase factors\\
Projection & $\pi:\mathcal{H}\backslash\left\{  0\right\}  \rightarrow
\mathcal{P}\left(  \mathcal{H}\right)  $ & Surjections\\\hline
\end{tabular}
\caption{Schematic summary of fiber bundle ingredients with corresponding
quantum mechanical quantities (symbols), along with their elements. For
instance, the space of rays $\mathcal{P}\left(  \mathcal{H}\right)  $ in
quantum mechanics corresponds to the base space in the fiber bundle
construction.}%
\end{table}

Clearly, a path $t\mapsto\left\vert \psi\left(  t\right)  \right\rangle $ with
$0\leq t\leq T$ traced out by a state vector $\left\vert \psi\left(  t\right)
\right\rangle $ satisfying the evolution equation $i\hslash\partial
_{t}\left\vert \psi\left(  t\right)  \right\rangle =\mathrm{H}\left(
t\right)  \left\vert \psi\left(  t\right)  \right\rangle $ lies in
$\mathcal{H}\backslash\left\{  0\right\}  $. The corresponding path in
$\mathcal{P}\left(  \mathcal{H}\right)  $ can be determined by projecting the
path in $\mathcal{H}\backslash\left\{  0\right\}  $ down onto a path in
$\mathcal{P}\left(  \mathcal{H}\right)  $. It is interesting to point out that
in a cyclic quantum evolution, the initial and final physical states are the
same. Therefore, cyclic evolutions are closed paths in $\mathcal{P}\left(
\mathcal{H}\right)  $. However, closed paths in $\mathcal{P}\left(
\mathcal{H}\right)  $ can correspond to open paths in $\mathcal{H}%
\backslash\left\{  0\right\}  $. Therefore, the initial and final state
vectors in a cyclic evolution are on the same fiber, but at different
\textquotedblleft heights\textquotedblright. Heights are characterized by the
emergence of an overall phase factor between the state vectors of interest,
$\left\vert \psi\left(  T\right)  \right\rangle =e^{i\phi_{\mathrm{tot}%
}\left(  T\right)  }\left\vert \psi\left(  0\right)  \right\rangle $. Part of
the total phase $\phi_{\mathrm{tot}}\left(  T\right)  $ (that is, the
geometric phase $\phi_{\mathrm{geometric}}\left(  T\right)  $) depends only on
the geometry of the fiber bundle.

The geometry of the fiber bundle is characterized by a connection that helps
comparing fibers at different points on $\mathcal{P}\left(  \mathcal{H}%
\right)  $. The connection, in turn, can be introduced once one considers the
decomposition of the tangent space \textrm{T}$\left[  \mathcal{H}%
\backslash\left\{  0\right\}  \right]  $ to $\mathcal{H}\backslash\left\{
0\right\}  $ in terms of an horizontal space \textrm{H}$_{\mathrm{T}\left[
\mathcal{H}\backslash\left\{  0\right\}  \right]  }$ and a vertical subspace
\textrm{V}$_{\mathrm{T}\left[  \mathcal{H}\backslash\left\{  0\right\}
\right]  }$,%
\begin{equation}
\mathrm{T}\left[  \mathcal{H}\backslash\left\{  0\right\}  \right]
=\mathrm{V}_{\mathrm{T}\left[  \mathcal{H}\backslash\left\{  0\right\}
\right]  }\oplus\mathrm{H}_{\mathrm{T}\left[  \mathcal{H}\backslash\left\{
0\right\}  \right]  }\text{.} \label{deco}%
\end{equation}
In terms of Eq. (\ref{deco}), \ a vector $\left\vert \dot{\psi}\left(
t\right)  \right\rangle \in\mathrm{T}\left[  \mathcal{H}\backslash\left\{
0\right\}  \right]  $ with $\dot{\psi}\overset{\text{def}}{=}d\psi/dt$ can be
decomposed as%
\begin{equation}
\left\vert \dot{\psi}\left(  t\right)  \right\rangle =\left\langle \psi\left(
t\right)  |\dot{\psi}\left(  t\right)  \right\rangle \left\vert \psi\left(
t\right)  \right\rangle +\left\vert h_{\psi}\left(  t\right)  \right\rangle
\text{,}%
\end{equation}
where $\left\vert h_{\psi}\left(  t\right)  \right\rangle \overset{\text{def}%
}{=}\left[  \left\vert \dot{\psi}\left(  t\right)  \right\rangle -\left\langle
\psi\left(  t\right)  |\dot{\psi}\left(  t\right)  \right\rangle \left\vert
\psi\left(  t\right)  \right\rangle \right]  \perp\left\vert \psi\left(
t\right)  \right\rangle $, $\left\vert h_{\psi}\left(  t\right)  \right\rangle
\in\mathrm{H}_{\mathrm{T}\left[  \mathcal{H}\backslash\left\{  0\right\}
\right]  }$, and $\left\langle \psi\left(  t\right)  |\dot{\psi}\left(
t\right)  \right\rangle \left\vert \psi\left(  t\right)  \right\rangle
\in\mathrm{V}_{\mathrm{T}\left[  \mathcal{H}\backslash\left\{  0\right\}
\right]  }$. Observe that the connection one-form $\mathcal{A}\left(
t\right)  \overset{\text{def}}{=}-i\left\langle \psi\left(  t\right)
\left\vert \dot{\psi}\left(  t\right)  \right.  \right\rangle $\textbf{
}appears in the definition of $\left\vert h_{\psi}\left(  t\right)
\right\rangle $, the covariant derivative of $\left\vert \psi\left(  t\right)
\right\rangle $\textbf{. }Indeed, consider the projector\textbf{
}$P_{\parallel}\overset{\text{def}}{=}\left\vert \psi\right\rangle
\left\langle \psi\right\vert $ onto $\left\vert \psi\right\rangle $ and the
projector $P_{\perp}=\mathrm{I}-\left\vert \psi\right\rangle \left\langle
\psi\right\vert $ onto states perpendicular to $\left\vert \psi\right\rangle
$. Then, $P_{\parallel}+P_{\perp}=\mathrm{I}$ is a resolution of the identity
operator $\mathrm{I}$, $P_{\parallel}^{2}=P_{\parallel}$, $P_{\perp}%
^{2}=P_{\perp}$, and $P_{\parallel}P_{\perp}=P_{\perp}P_{\parallel}%
=\mathrm{O}$ is the null operator. Using\textbf{ }the definitions of
$\mathcal{A}\left(  t\right)  $, $P_{\parallel}$, and\textbf{ }$P_{\perp}$, we
find $\left\vert h_{\psi}\left(  t\right)  \right\rangle =\left\vert \dot
{\psi}\left(  t\right)  \right\rangle -i\mathcal{A}\left(  t\right)
\left\vert \psi\left(  t\right)  \right\rangle $ \cite{samuel88}\textbf{. }For
further details on this construction, \textbf{see} Ref. \cite{bohm91}.
Finally, once a connection is specified, the concept of a horizontal lift can
be properly defined. In particular, a horizontal lift is specified by lifting
the tangent vectors of a curve in $\mathcal{P}\left(  \mathcal{H}\right)  $ to
horizontal tangent vectors of a curve in $\mathcal{H}\backslash\left\{
0\right\}  $. We are now ready to introduce the concept of Berry's geometric phase.

\emph{Remarks on the geometric phase}. In June of 1983, considering an
adiabatic (i.e., slow varying parameters) and cyclic quantum evolution of a
quantum state $\left\vert \psi\left(  t\right)  \right\rangle $ as an
eigenstate of the Hamiltonian \textrm{H}$\left(  t\right)  $ of the system in
a time interval $T$, Michael Berry discovered that the quantum state gains a
geometrical phase factor in addition to the usual dynamical phase factor
\cite{berry84}. The original state $\left\vert \psi\left(  0\right)
\right\rangle $\ returns to itself up to a phase factor,%
\begin{equation}
\left\vert \psi\left(  T\right)  \right\rangle =e^{i\phi_{\mathrm{tot}}\left(
T\right)  }\left\vert \psi\left(  0\right)  \right\rangle \text{.}
\label{giro}%
\end{equation}
The total phase $\phi_{\mathrm{tot}}\left(  T\right)  $ in Eq. (\ref{giro}) is
the sum of the dynamical phase $\phi_{\mathrm{dynamical}}\left(  T\right)  $
and the geometric phase $\phi_{\mathrm{geometric}}\left(  T\right)  $,
\begin{equation}
\phi_{\mathrm{tot}}\left(  T\right)  =\phi_{\mathrm{dynamical}}\left(
T\right)  +\phi_{\mathrm{geometric}}\left(  T\right)  \text{,} \label{totf}%
\end{equation}
with $\phi_{\mathrm{dynamical}}\left(  T\right)  $ defined as%
\begin{equation}
\phi_{\mathrm{dynamical}}\left(  T\right)  \overset{\text{def}}{=}-\int%
_{0}^{T}\frac{\left\langle \psi\left(  t\right)  |\mathrm{H}\left(  t\right)
|\psi\left(  t\right)  \right\rangle }{\left\langle \psi\left(  t\right)
|\psi\left(  t\right)  \right\rangle }dt\text{.} \label{df}%
\end{equation}
In the fiber bundle description of quantum mechanics \cite{eguchi80}, one can
regard the space of normalized states as a fiber bundle over the space of
rays, with the bundle having a natural connection that allows to compare the
phases on two neighboring states. In October of 1983, using this fiber bundle
formalism, Barry Simon interpreted in \cite{simon83} the geometric phase
$\phi_{\mathrm{geometric}}\left(  T\right)  \overset{\text{def}}{=}%
\phi_{\mathrm{tot}}\left(  T\right)  -\phi_{\mathrm{dynamical}}\left(
T\right)  $ as the line integral of the Abelian connection one-form $A$ (for
the geometrical phase on $\mathcal{P}\left(  \mathcal{H}\right)  $) over a
closed path $l$ in the projective Hilbert space $\mathcal{P}\left(
\mathcal{H}\right)  $,%
\begin{equation}
\phi_{\mathrm{geometric}}\left(  T\right)  =%
%TCIMACRO{\doint \limits_{l}}%
%BeginExpansion
{\displaystyle\oint\limits_{l}}
%EndExpansion
\mathcal{A}=\int_{\Sigma}\mathcal{F}\text{.} \label{Berry}%
\end{equation}
The second equality in Eq. (\ref{Berry}) is a consequence of Stokes' theorem
(i.e., recasting a line integral as a surface integral), with $\mathcal{F}%
\overset{\text{def}}{=}D\mathcal{A}$ denoting the Abelian curvature two-form
and $\Sigma$ being any surface bounded by $l$ in $\mathcal{P}\left(
\mathcal{H}\right)  $. The geometrical nature of this phase in Eq.
(\ref{Berry}) is justified by its dependence solely on the closed path
evolution of the ray in the projective Hilbert space and, moreover, by its
complete independence on any aspect of the Hamiltonian that governs the
dynamical evolution. In 1987, Aharonov and Anandan showed in \cite{aa87} that
the adiabaticity requirement is unnecessary for the emergence of geometric
phases in cyclic quantum evolutions. In 1988, Samuel and Bhandari introduced
in \cite{samuel88} a general setting for Berry's geometric phase in which
neither unitarity nor cyclicity of the quantum evolution are required. In a
series of works between 1991 and 1995 \cite{pati91,pati94,pati95}, Pati
devoted a serious effort in describing the relation between phases and
distances in quantum evolutions, both cyclic \cite{pati94} and noncyclic
\cite{pati95}. In \cite{pati95}, Pati expressed the geometric phase factor
$e^{i\phi_{\mathrm{geometric}}\left(  T\right)  }$ in a noncyclic evolution
with $0\leq t\leq T$ in terms of the horizontal lift $\left\vert \bar{\psi
}\left(  t\right)  \right\rangle $ of a curve in the projective Hilbert space
$\mathcal{P}\left(  \mathcal{H}\right)  $ as%
\begin{equation}
e^{i\phi_{\mathrm{geometric}}\left(  T\right)  }\overset{\text{def}}{=}%
\frac{\left\langle \bar{\psi}\left(  0\right)  |\bar{\psi}\left(  T\right)
\right\rangle }{\left\vert \left\langle \bar{\psi}\left(  0\right)  |\bar
{\psi}\left(  T\right)  \right\rangle \right\vert }\text{. } \label{triangolo}%
\end{equation}
The horizontal lift $\left\vert \bar{\psi}\left(  t\right)  \right\rangle $ is
defined in terms of the state $\left\vert \psi\left(  t\right)  \right\rangle
$ that satisfies $i\hslash\partial_{t}\left\vert \psi\left(  t\right)
\right\rangle =\mathrm{H}\left(  t\right)  \left\vert \psi\left(  t\right)
\right\rangle $ as \cite{pati95},%
\begin{equation}
\left\vert \bar{\psi}\left(  t\right)  \right\rangle \overset{\text{def}%
}{=}e^{\frac{i}{\hslash}\int_{0}^{t}\left\langle \psi\left(  t^{\prime
}\right)  |\mathrm{H}\left(  t^{\prime}\right)  |\psi\left(  t^{\prime
}\right)  \right\rangle dt^{\prime}}\left\vert \psi\left(  t\right)
\right\rangle \text{,} \label{horizontal}%
\end{equation}
with $\left\langle \bar{\psi}\left(  t\right)  |\partial_{t}\bar{\psi}\left(
t\right)  \right\rangle =0$. Interestingly, observe that the phase factor in
the horizontal lift\textbf{ }$\left\vert \bar{\psi}\left(  t\right)
\right\rangle $\textbf{ }in Eq. (\ref{horizontal}) can be\textbf{ }expressed
in terms of the connection\textbf{ }$\mathcal{A}\left(  t\right)  $\textbf{.
}Indeed, since\textbf{ }$i\hslash\partial_{t}\left\vert \psi\left(  t\right)
\right\rangle =\mathrm{H}\left(  t\right)  \left\vert \psi\left(  t\right)
\right\rangle $\textbf{ }and\textbf{ }$\left\langle \psi\left(  t\right)
\left\vert \dot{\psi}\left(  t\right)  \right.  \right\rangle =i\mathcal{A}%
\left(  t\right)  $\textbf{, }we have\textbf{ }$i\left\langle \psi\left(
t\right)  \left\vert \mathrm{H}\right\vert \psi\left(  t\right)  \right\rangle
=-i\hslash\mathcal{A}\left(  t\right)  $\textbf{. }Therefore, Eq.
(\ref{horizontal}) can be recast as%
\begin{equation}
\left\vert \bar{\psi}\left(  t\right)  \right\rangle \overset{\text{def}%
}{=}e^{-\frac{i}{\hslash}\int_{0}^{t}\mathcal{A}\left(  t^{\prime}\right)
dt^{\prime}}\left\vert \psi\left(  t\right)  \right\rangle \text{.}%
\end{equation}
In 1993, Mukunda and Simon provided in \cite{mukunda93} a very general setting
for the geometric phase for any smooth open curve of unit vectors in Hilbert
space by employing the kinematics of the Hilbert space of states of a general
quantum system along with a properly defined gauge transformation group. In
particular, they showed that the total phase along the horizontal lift of a
geodesic curve in $\mathcal{P}\left(  \mathcal{H}\right)  $ is zero.
Therefore, since the dynamical phase along an horizontal lift is zero, they
concluded that geodesics are null (geometric) phase curves. In the geodesic
curve scenario, $\left\langle \bar{\psi}\left(  0\right)  |\bar{\psi}\left(
T\right)  \right\rangle $ is real and positive. Therefore, $\left\langle
\bar{\psi}\left(  0\right)  |\bar{\psi}\left(  T\right)  \right\rangle
=\left\vert \left\langle \bar{\psi}\left(  0\right)  |\bar{\psi}\left(
T\right)  \right\rangle \right\vert $ in Eq. (\ref{triangolo}), $e^{i\phi
_{\mathrm{geometric}}\left(  T\right)  }=1$ and, finally, $\phi
_{\mathrm{geometric}}\left(  T\right)  =0$. In conclusion,
\begin{equation}
\phi_{\mathrm{geometric}}^{\left(  \gamma_{\mathrm{geodesic}}\right)  }\left(
T\right)  =0 \label{nullphase}%
\end{equation}
is the necessary but not sufficient condition for geodesic behavior of a curve
$\gamma_{\mathrm{geodesic}}$ in $\mathcal{P}\left(  \mathcal{H}\right)  $.
Before presenting this verification, we present for completeness a quick
remark. As pointed out earlier, we stated there exist curves connecting two
pure state that are not necessarily the shortest curves for which, however,
the gained geometric phase is zero. These curves generalize the concept of
geodesic curve and are known in the literature as null phase curves (NPCs,
\cite{rabei99,mukunda03,chaturvedi13}). A geodesic is a NPC. The converse, in
general, is false. The generalization involved in transitioning from geodesics
to NPCs emerges especially when the dimensionality of the complex Hilbert
space is greater than or equal to three. Indeed, in a two-dimensional Hilbert
space, the ray space is the Poincar\'{e} sphere and NPCs are great circles
arcs or geodesics on the sphere $\mathcal{S}^{2}$. A NPC connecting any two
nonantipodal points on $\mathcal{S}^{2}$ is either the corresponding geodesic,
or it may navigate some extended region of the corresponding great circle.
When $\dim_{%
%TCIMACRO{\U{2102} }%
%BeginExpansion
\mathbb{C}
%EndExpansion
}\mathcal{H}>2$, NPCs are more numerous than geodesics and there are
infinitely many NPCs connecting any two nonorthogonal points in ray space
(against a single geodesic). Examples of NPCs that are not (free) geodesics
can be found in the framework of the so-called constrained geodesics
\cite{rabei99}. Unlike free geodesics, constrained geodesics are paths of
minimum length connecting pairs of points on a smooth submanifold of the
complete ray space that specifies the physical system under consideration. Two
illustrative physical examples of such curves are constrained geodesics on the
submanifolds of single mode coherent states and normalized Gaussian pure
states. In both cases, the constrained geodesics differ from the free
geodesics. In the Gaussian case, for instance, constrained geodesic paths are
assumed to be traversing solely centered normalized Gaussian wave functions
and not even superpositions of Gaussians are to be considered. However, in
both examples, it happens that constrained geodesics are also null phase
curves. For more details, we refer to Ref. \cite{rabei99}. For a formal
definition of a NPC in terms of a real and positive Bargmann invariant of
third order, we refer to Ref. \cite{mukunda03}. We now return to the verification.

\emph{The verification}. We note that the horizontal lift $\left\vert
\bar{\psi}\left(  t\right)  \right\rangle $ in\ Eq. (\ref{horizontal}) equals
$\left\vert \psi\left(  t\right)  \right\rangle $ in Eq. (\ref{geodesic}).
Indeed, in our problem\textbf{, }$\left\langle \psi\left(  t\right)
|\mathrm{H}\left(  t\right)  |\psi\left(  t\right)  \right\rangle =0$ with
$\mathrm{H}\left(  t\right)  $ in Eq. (\ref{amy})\textbf{.} Alternatively,
using Eq. (\ref{geodesic}), one can use brute force to verify that
$\left\langle \psi\left(  t\right)  |\partial_{t}\psi\left(  t\right)
\right\rangle =i\mathcal{A}\left(  t\right)  =0$. We also observe that
$T\overset{\text{def}}{=}\left(  \hslash/E\right)  \left[  \left(
\varphi_{\alpha}-\varphi_{\beta}\right)  /2\right]  $, $\left\vert
\psi(0)\right\rangle \overset{\text{def}}{=}\left\vert A\right\rangle $ in Eq.
(\ref{chi5a}), and $\left\vert \psi\left(  T\right)  \right\rangle
\overset{\text{def}}{=}e^{i\frac{\varphi_{\alpha}-\varphi_{\beta}}{2}%
}\left\vert B\right\rangle \neq\left\vert A\right\rangle $ with $\left\vert
B\right\rangle $ defined in Eq. (\ref{chi5b}). Therefore, recalling that
$\left\langle A|B\right\rangle \overset{\text{def}}{=}e^{-i\frac
{\varphi_{\alpha}-\varphi_{\beta}}{2}}\cos\left[  \left(  \varphi_{\alpha
}-\varphi_{\beta}\right)  /2\right]  $, we get $\left\langle A|\psi\left(
T\right)  \right\rangle =\cos\left(  \theta_{\mathrm{FS}}/2\right)  \in%
%TCIMACRO{\U{211d} }%
%BeginExpansion
\mathbb{R}
%EndExpansion
_{+}$ with $\theta_{\mathrm{FS}}\overset{\text{def}}{=}\varphi_{\alpha
}-\varphi_{\beta}=2s_{\mathrm{FS}}=s_{\mathrm{geo}}\leq\pi$. Clearly,
$s_{\mathrm{FS}}$ and $s_{\mathrm{geo}}$ denote the Fubini-Study and the
geodesic distances, respectively. Finally, employing Eq. (\ref{triangolo}), we
conclude that Eq. (\ref{nullphase}) is properly fulfilled.

\section{Concluding Remarks}

In this paper, we presented an explicit geodesic analysis of the dynamical
trajectories that emerge from the quantum evolution of a single-qubit quantum
state. The evolution is governed by an Hermitian Hamiltonian operator that
leads to the fastest possible unitary evolution between given initial and
final states. \ To achieve maximum clarity, we proceeded in a step-by-step
fashion. First, we reviewed preliminary material on the geometry of pure
quantum states. In particular, we emphasized the notions of quantum lines
(Eqs. (\ref{geo11}) and (\ref{geo12})) and quantum geodesic lines (Eq.
(\ref{ass2})) on a curved manifold of pure states equipped with the
Fubini-Study metric (Eq. (\ref{FS1})). Second, we introduced an optimal-speed
Hamiltonian evolution (Eq. (\ref{amy})) and calculated the corresponding
shortest time quantum dynamical trajectory traced by the evolved quantum state
(Eq. (\ref{geodesic})). Finally, combining facts and results discussed in the
first two steps, we explicitly checked in the final third step the geodesicity
of the quantum dynamical trajectory emerging from the chosen optimal-speed
Hamiltonian evolution via the geodesicity condition in Eq. (\ref{ass2}). The
key observation in this derivation was the clever sequential change of
parametrizations (Eqs. (\ref{geodesic}), (\ref{GG23}), (\ref{para2}) and
Figure 1 along with Table I) that leads, finally, to the geodesic path on the
Bloch sphere. For completeness, we also studied the geodesic behavior of the
quantum trajectory in terms of the concepts of geometric efficiency in Eq.
(\ref{efficiency}) and of Berry's geometric phase in Eq. (\ref{Berry}).
Working out in detail the optimal-speed Hamiltonian evolution of a two-level
quantum system specified in Eq. (\ref{amy}), we provided three alternative
perspectives on the geodesic behavior of a curve $\gamma_{A\rightarrow
B}\left(  t\right)  $ with $0\leq t\leq T$ in ray space $\mathcal{P}\left(
\mathcal{H}\right)  $ (with $\mathcal{H=H}_{2}^{1}$ and $\mathcal{P}\left(
\mathcal{H}_{2}^{1}\right)  =%
%TCIMACRO{\U{2102} }%
%BeginExpansion
\mathbb{C}
%EndExpansion
P^{1}\cong S^{2}$):

\begin{description}
\item[{[i]}] Geodesics can be viewed as minimal length curves (Eq.
(\ref{geocondition})),%
\begin{equation}
\mathrm{Length}\left(  \gamma_{A\rightarrow B}^{\left(  \mathrm{geodesic}%
\right)  }\right)  \leq\mathrm{Length}\left(  \gamma_{A\rightarrow B}^{\left(
\mathrm{non}\text{-}\mathrm{geodesic}\right)  }\right)  \text{.} \label{A}%
\end{equation}

\item[{[ii]}] Geodesics can be regarded as unit geometric efficiency curves
(Eq. (\ref{unitefficiency})),%
\begin{equation}
\eta_{\mathrm{geometric}}^{\left(  \gamma_{\mathrm{geodesic}}\right)
}=1\text{.} \label{B}%
\end{equation}

\item[{[iii]}] Geodesics can be considered as null phase curves (Eq.
(\ref{nullphase})),%
\begin{equation}
\phi_{\mathrm{geometric}}^{\left(  \gamma_{\mathrm{geodesic}}\right)  }\left(
T\right)  =0\text{.} \label{C1}%
\end{equation}

\end{description}

As pointed out in the paper, the relations in Eqs. (\ref{A}) and (\ref{B})
provide necessary and sufficient geodesicity conditions. The relation in Eq.
(\ref{C1}), instead, is only a necessary condition.

From a quantum mechanics standpoint, we focused in this paper on the study of
the time-optimal evolution of closed two-level quantum systems specified by
pure states driven by the Schr\"{o}dinger equation. From a geometry viewpoint,
we used the Fubini-Study distance measure since it is the only natural choice
for a measure that defines \textquotedblleft random states\textquotedblright.
Specifically, the distance between density matrices for both pure and mixed
states must decrease under coarse-graining (i.e., randomization), if the
distance expresses statistical distinguishability \cite{petz96a,petz99}. In
this respect, the Fubini-Study metric is the only monotone Riemannian metric
on the space of pure quantum states. However, a more realistic scenario is the
case of open system dynamics in mixed quantum states
\cite{taddei13,adolfo13,deffner13}. In this case, a number of new challenges
are expected to emerge. First, one needs to consider general nonunitary
quantum evolutions where the dynamics is described by a master equation in the
Lindblad form. Finding exact analytical expressions for the actual dynamical
trajectories traced by an open quantum system in a mixed quantum state can be
rather complicated \cite{carlini08,brody19}. For an explicit discussion on
some conceptual and computational difficulties in finding the time-optimal
quantum evolution of mixed states governed by a master equation, we refer to
Ref. \cite{carlini08}. Indeed, even limiting the attention to the unitary
evolution of closed physical systems, optimal-time evolutions of mixed (pure)
states are typically generated by time-varying (constant) Hamiltonians
\cite{hornedal22}. Second, from a geometric perspective, there are infinitely
many monotone Riemannian metrics on the space of mixed quantum states as
specified by the Morozova-Cencov-Petz theorem \cite{petz96a,petz99}.

Therefore, there is the freedom to choose a variety of distance measures
between mixed states. There is the need to study several measures, each of
them with specific physical motivations, convenience, disadvantage. Arguably,
the main hurdle when investigating open systems in mixed quantum states is
this nonuniqueness of the metric. However, even assuming to have chosen the
metric and having ready to use the actual dynamical trajectory of the system,
it is generally not straightforward finding closed form expressions of
geodesic paths on arbitrary manifolds of mixed quantum states equipped with
Riemann metrics of statistical relevance. In a few cases, however, this task
can be successfully\textbf{ }accomplished \cite{weis13}. For example, geodesic
paths connecting two mixed states are known for some metrics, including the
Quantum Fisher information (QFI) metric \cite{uhlmann95}, the Wigner-Yanase
metric \cite{gibilisco03}, and the metric based on the trace distance
\cite{cai17}. In particular, in the case of the Bures metric
\cite{bures69,uhlmann76,hubner92} (or, alternatively, QFI metric with\textbf{
}$g_{\mu\nu}^{\left(  \mathrm{QFI}\right)  }=4g_{\mu\nu}^{\left(
\mathrm{Bures}\right)  }$\textbf{), }formulas for geodesic paths can be
presented in terms of projections of large circles on a sphere in a purifying
space \cite{dittman95}. Furthermore, in the Wigner-Yanase metric case
\cite{wigner63,luo03}, it is possible to provide explicit expressions for the
geodesic path, geodesic distance, and, finally, sectional and scalar
curvatures. These relations were originally determined by Gibilisco and Isola
in Ref. \cite{gibilisco03} by mimicking from a quantum-mechanical perspective
the classical pull-back approach to the Fisher information.

Being aware of the above mentioned challenges, our analysis presented in this
paper could be potentially extended in several ways. \ First, keeping the
dimensionality of the Hilbert space fixed at two, we could build on our
findings to pursue a comparative analysis of geodesic paths of qubits in mixed
quantum states inside the Bloch sphere equipped with different Riemannian
metrics \cite{erik20,cafaroprd22}. Second, deviating from the condition of
optimal-speed quantum Hamiltonian evolutions \cite{carlo19A}, we could rely on
our work to characterize nearly-optimal quantum evolutions in\textbf{
}geometric terms by focusing on variations from geodesics and, moreover, on
curvature effects between nearby geodesic paths
\cite{cafaro07,cafaro11,laba17}. Third, as the dimension of the Hilbert space
increases, we expect the geometry of the state spaces to become richer with a
very intricate structure. New features can arise in higher-dimensional Hilbert
spaces, including multipartite quantum entanglement. A first step in this
higher-dimensional setting could be that of considering the geometry of the
Bloch vector\textbf{ }for\textbf{ }$d$\textbf{-}level quantum systems\textbf{
}(i.e., qudits) with\textbf{ }$d>2$ \cite{bertlmann08}. In particular, it
would seem reasonable to begin by studying the geometry of a qutrit (i.e., a
three-level quantum system) specified by\textbf{ }$d=3$\textbf{ }%
\cite{kurzynski16,goyal16}.

In conclusion, despite its limited focus on dimension $d=2$, we hope our work
will stimulate other researchers and open the way toward further explicit
investigations on the interplay between quantum mechanics and geometry. For
the time being, we leave a more in-depth quantitative discussion on these
potential geometric extensions of our analytical findings, including
generalizations to mixed state geometry and quantum evolutions, to future
scientific efforts.

\begin{acknowledgments}
C.C. is grateful to the United States Air Force Research Laboratory (AFRL)
Summer Faculty Fellowship Program for providing support for this work. P.M.A.
acknowledges support from the Air Force Office of Scientific Research (AFOSR).
Any opinions, findings and conclusions or recommendations expressed in this
material are those of the author(s) and do not necessarily reflect the views
of the Air Force Research Laboratory (AFRL). The authors thank an anonymous
referee for very useful comments leading to an improved version of this manuscript.
\end{acknowledgments}

\end{document}